\documentclass[twocolumn]{aastex631}

\usepackage{comment}
\usepackage[normalem]{ulem}
\usepackage{amsmath}
\usepackage{orcidlink}
\usepackage{graphicx}
\usepackage{array}
\newcolumntype{P}[1]{>{\centering\arraybackslash}p{#1}}

\graphicspath{{figures/}}

\newcommand{\llnl}{Lawrence Livermore National Laboratory, 7000 East Ave., Livermore, California 94609 USA}
\newcommand{\ucr}{Department of Physics and Astronomy, University of California, Riverside, CA 92521, USA}
\newcommand{\ucb}{Department of Astronomy, University of California, Berkeley, Berkeley, CA 94720, USA}
\newcommand{\carnegie}{Observatories of the Carnegie Institution for Science, Pasadena, CA 91101, USA}

\newcommand{\loc}{\mu}

\newcommand{\eparam}{\boldsymbol{\theta}}
\newcommand{\eparaml}{\boldsymbol{\theta}_l}
\newcommand{\eparamc}{\boldsymbol{\theta}_c}
\newcommand{\eparamdi}{\boldsymbol{\theta}_i}
\newcommand{\eparamndj}{\bar{\boldsymbol{\theta}}_j}

\newcommand{\eparamnl}{\boldsymbol{\vartheta}}

\newcommand{\D}{\boldsymbol{d}}

\newcommand{\Dcj}{\boldsymbol{d}_{c,j}}
\newcommand{\Dt}{d_t}

\newcommand{\DDi}{\boldsymbol{d}_i}
\newcommand{\DNDj}{\bar{\boldsymbol{d}}_j}
\newcommand{\DNEk}{\bar{\bar{\boldsymbol{d}}}_k}
\newcommand{\classa}{{\rm class}_a}
\newcommand{\classb}{{\rm class}_b}
\newcommand{\classBH}{{\rm class}_{\rm BH}}
\newcommand{\nclass}{N_{\rm pop} }

\newcommand{\hypshapea}{\boldsymbol{\lambda_a}}

\newcommand{\hyp}{\boldsymbol{\Lambda} }
\newcommand{\NOBS}{N_{\rm obs} }
\newcommand{\NNOBS}{N_{\neg {\rm obs}} }
\newcommand{\NNOEVENT}{N_{\emptyset} }
\newcommand{\Nt}{N_{T} }
\newcommand{\trig}{{\rm tr} }
\newcommand{\trigdi}{{\rm tr}_i }
\newcommand{\trigndj}{\bar{{\rm tr}}_j }
\newcommand{\trignek}{\bar{\bar{{\rm tr}}}_k }

\newcommand{\popfrac}{\boldsymbol{\psi} }
\newcommand{\popfraca}{\psi_a }

\newcommand{\popA}{{\rm Star}}
\newcommand{\popB}{{\rm SOBH}}
\newcommand{\popC}{{\rm PBH}}
\newcommand{\fracA}{\psi_{\rm Star}}
\newcommand{\fracB}{\psi_{\rm SOBH}}
\newcommand{\fracC}{\psi_{\rm PBH}}

\newcommand{\scale}{\sigma }

\newcommand{\IMRELEASENO}{LLNL-JRNL-852673-DRAFT}

\begin{document}

\title{Disentangling the Black Hole Mass Spectrum with Photometric Microlensing Surveys}

\begin{abstract}

\noindent From the formation mechanisms of stars and compact objects to nuclear physics, modern astronomy frequently leverages surveys to understand populations of objects to answer fundamental questions. The population of dark and isolated compact objects in the Galaxy contains critical information related to many of these topics, but is only practically accessible via gravitational microlensing. However, photometric microlensing observables are degenerate for different types of lenses, and one can seldom classify an event as involving either a compact object or stellar lens on its own. To address this difficulty, we apply a Bayesian framework that treats lens type probabilistically and jointly with a lens population model. This method allows lens population characteristics to be inferred despite intrinsic uncertainty in the lens-class of any single event. We investigate this method's effectiveness on a simulated ground-based photometric survey in the context of characterizing a hypothetical population of primordial black holes (PBHs) with an average mass of $30 M_{\odot}$. On simulated data, our method outperforms current black hole (BH) lens identification pipelines and characterizes different subpopulations of lenses while jointly constraining the PBH contribution to dark matter to ${\approx}25$\%. Key to robust inference, our method can marginalize over population model uncertainty. We find the lower mass cutoff for stellar origin BHs, a key observable in understanding the BH mass gap, particularly difficult to infer in our simulations. This work lays the foundation for cutting-edge PBH abundance constraints to be extracted from current photometric microlensing surveys.

\end{abstract}

\author[0000-0002-5910-3114]{Scott E. Perkins\,}
\email{perkins35@llnl.gov}
\affiliation{\llnl}
\author[0000-0002-1052-6749]{Peter McGill\,}
\affiliation{\llnl}
\author[0000-0003-0248-6123]{William Dawson\,}
\affiliation{\llnl}
\author[0000-0002-0287-3783]{Natasha S. Abrams\,}
\affiliation{\ucb}
\author[0000-0002-6406-1924]{Casey Y. Lam\,}
\affiliation{\ucb}
\affiliation{\carnegie}
\author[0000-0002-4457-890X]{Ming-Feng Ho\,}
\affiliation{\ucr}

\author[0000-0001-9611-0009]{Jessica R.~Lu\,}
\affiliation{\ucb}
\author[0000-0001-5803-5490]{Simeon Bird\,}
\affiliation{\ucr}
\author[0000-0002-2911-8657]{Kerianne Pruett}
\affiliation{\llnl}
\author[0000-0003-2632-572X]{Nathan Golovich}
\affiliation{\llnl}
\author{George Chapline}
\affiliation{\llnl}

\section{Introduction}\label{sec:intro}


Understanding the population of black holes (BHs) in our galaxy and universe will shed light on outstanding problems in both cosmology \citep[e.g.,][]{Zeldovich1967,Clesse:2015wea,1974A&A....37..225M,Farrah:2023opk} and astrophysics \citep[e.g.,][]{2001ApJ...554..548F,2010ApJ...725.1918O,2011ApJ...741..103F,Bouffanais:2019nrw,Baibhav:2020xdf,Mandel:2018hfr}.
Of particular interest for both of these topics is the investigation of different formation channels of BHs, both astrophysical \citep[e.g.,][]{2003ApJ...591..288H,2016ApJ...818..124E,2019BAAS...51c.365L,2021MNRAS.501.4514C} and cosmological \citep[e.g.,][]{Car1974, 1974A&A....37..225M,Garcia-Bellido:1996mdl,PhysRevD.12.2949}.
The latter of these groups of formation channels produces primordial BHs \citep[PBHs;][]{Zeldovich1967}.
If observed, a population of PBHs could explain some fraction of dark matter \citep{Carr:2016drx,Bird2023} and the first supermassive BH seeds \citep{Kawasaki:2012kn,Inayoshi:2019fun,Bean:2002kx,Rice:2017avg,Carr:2018rid}, providing a unique probe into the early universe \citep{Car1975,1975Natur.253..251C}.

Unfortunately, BHs in isolation do not radiate measurable amounts of light, gravitational radiation or particles making them difficult to detect. Detectable emission from a BH is only produced through interaction with its environment. Massive, extra-galactic BHs can be detected through a strong gravitational interaction with another object causing gravitational radiation \citep[e.g.,][]{LIGOScientific:2016aoc,LIGOScientific:2021djp} or through accretion, causing electromagnetic (EM) radiation \citep[e.g.,][]{EventHorizonTelescope:2019dse,2022ApJ...930L..12E,Fabbiano:2005pj}.  
Studies using gravitational wave (GW) emission and EM observation can be effective for understanding the extra-galactic BH population \citep[e.g.,][]{KAGRA:2021duu,LIGOScientific:2020kqk,LIGOScientific:2021psn,Edelman:2022ydv,Roulet:2018jbe}, provided that detection bias from observational selection effects is mitigated \citep{Liotine:2022vwq}. 

Within the Milky Way, there are estimated to be $\approx10^{8}$ stellar origin BHs \citep[SOBHs; e.g.,][]{Samland1998}. Despite this large expected abundance, only ${\sim}50$ SOBHs have been detected. The bulk of these BHs are found in X-ray binaries \citep[e.g.,][]{Remillard2006,Corral-Santana2016}, despite these systems being an intrinsically rare outcome of binary evolution \citep[e.g.,][]{Kalogera2001, El-Badry2023}. These systems are detectable due to bright X-ray emission from accretion of a luminous stellar companion onto the BH. Most recently, leveraging high-precision astrometry from Gaia \citep{Gaia2016,GaiaDR32021}, two nearby BHs that perturb the motion of their luminous binary companion have also been detected \citep{El-Badry2023a,El-Badry2023,Chakrabarti2022}.

Despite this diverse set of observational channels, they all require the BH, regardless of SOBH or PBH origin, to have a companion. 
None of these techniques are sensitive to detecting the population of isolated BHs within the the Galaxy.
Gravitational microlensing is uniquely positioned to fill this detection blind spot, as it is the only practical method with which isolated BHs can be detected and characterized~\citep{Gould2000, Bennett:2001vh,2022ApJ...933L..23L,Sahu2022,Chapline:2016tya}. 
Detecting a BH via microlensing only requires its close alignment with a distance background star.
In addition to understanding BHs from single microlensing events, the characteristics of sets of microlensing events observed over the course of a survey can encode information about the underlying BH lens population \citep[e.g.,][]{Lam_2020,Rose:2022xdu,2021arXiv210713697M,2020A&A...636A..20W,2016MNRAS.458.3012W,2021AcA....71...89M}.

However, robustly characterising the underlying lens population from microlensing surveys is challenging. This is because the photometric microlensing signal is degenerate in lens mass, distance, kinematics \citep{Paczynski1996}, typical transient survey noise systematics \citep{Golovich:2020acu}, and contains no direct lens mass or lens identity information. A microlensing event can also have an astrometric signal \citep{Eddington1919,Walker1995, Hog1995, Miyamoto1995, Rybicki2018}, which can break the photometric degeneracies resulting in a direct measurement of the lens mass \citep[e.g.,][]{Lu2016, Sahu2017,Kains2017,Zurlo2018,2022ApJ...933L..23L,Sahu2022,2023MNRAS.520..259M}, and also differentiate lens subpopulations \citep[e.g.,][]{Belokurov2002, Lam_2020, Pruett2022}. However, currently there is no large sample of microlensing events with measured astrometry, which would be required to perform population inference, although this is set to change over the coming years \citep[e.g.,][]{Gaia2016,2015arXiv150303757S, Lam_2020,Sajadian2023,Lam2023}.

In the absence of astrometry, ${\sim}10^{4}$ photometric microlensing events have been detected over the past decades \citep[e.g.,][]{Udalski2015, KMTnet2016, Jeong2015, Husseiniova2021} which can be used to constrain the underlying lens populations. Despite the degeneracies in the photometric signal, progress has been made in understanding the lens populations in the tails of the mass distribution i.e., Free Floating Planets \citep{2017Natur.548..183M,2023arXiv230308280S} and SOBHs \citep[e.g.,][]{2021arXiv210713697M} which effect the tails of the photometric microlensing event timescale distribution. However, current methods require manually pre-selecting or classifying events based on event characteristics, for example, assuming a set of candidate events with the longest timescales \citep[e.g.,][]{Lu2016} and large parallax signals are caused by BHs \citep[e.g.,][]{2016MNRAS.458.3012W, Kaczmarek2022}. 

In the case of candidate BH lenses, auxiliary information can sometimes be used to constrain the identity of the lens. This information includes, baseline source astrometry \citep[e.g.,][]{2020A&A...636A..20W, Kaczmarek2022}, testing if the event is consistent with the lens being dark, and assuming some model of the Galaxy which pins down the relative lens-source distance and kinematics \citep[e.g.,][]{2016MNRAS.458.3012W, Kaczmarek2022}. However, conclusions about the lens identity are sensitive to unreliable source astrometry and distances, and assumptions about the location of the lens and source imposed by a given Galactic model \citep{2021AcA....71...89M}.  Overall, definitively classifying the lens for a single microlensing event is difficult and can bias resulting inferences about the underlying lens population. 

In this work we overcome the lens classification problem by extending the inference framework of \cite{Zevin:2020gbd} and \cite{Franciolini:2021tla} and applying it to microlensing by treating the lens classification probabilistically. This method allows for all events to have some probability of belonging to each class (e.g, SOBH, PBH, or Star), effectively marginalizing over each possibility and bypassing the need to assume a single lens class. This approach allows the underlying lens population to be modelled jointly and in the absence of confident, individual event classifications. Our method generalizes the work of \cite{2023arXiv230308280S} and \cite{2021arXiv210713697M} to comprehensively include survey selection effects, rate information and probabilistic lens classification jointly with an uncertain lens population model. 

We apply this new framework with the goal of constraining and disentangling the mass spectra of the underlying lens population given a survey of photometric microlensing events. In this context, we focus on investigating if current photometric microlensing data can place constraints on the mass spectrum and abundance of PBHs in the Galaxy. Using simulated microlensing survey data, we evaluate the effectiveness of our method, including its ability to classify single lenses and marginalize over those classes to place constraints on the underlying lens populations. Through these exercises, we demonstrate the power of these methods on disentangling and constraining the mass spectra of PBHs and SOBHs and lay the foundation for these methods to be used in combination with Galactic simulations \citep[e.g.,][]{Lam_2020} to provide cutting edge constraints on the population of BHs in the Milky Way on current microlensing surveys like the Optical Gravitational Lensing Experiment \citep[OGLE;][]{Udalski2015}.

This work also complements methods applied decades ago to study MAssive Compact Halo Objects (MACHOs) by the MACHO \citep[e.g.,][]{Macho:2000nvd}, Expérience pour la Recherche d'Objets Sombres \citep[EROS;][]{EROS-2:2006ryy,2022A&A...664A.106B} and OGLE \citep[e.g.,][]{2009MNRAS.397.1228W,2010MNRAS.407..189W,2011MNRAS.413..493W,2011MNRAS.416.2949W} collaborations. These projects all used photometric microlensing observations of the Magellanic Clouds to estimate the abundance and halo mass fraction due to MACHOs (of which, PBHs could be a specific realization). These experiments probed the galactic halo, conservatively attributing all microlensing detections to MACHOs and using optical depth calculations. The methods proposed here are designed for use with observations of the galactic bulge, which observe thousands of events and must be understood in the context of many lensing subpopulations, requiring different statistical methods.

We begin by describing modeling of single photometric microlensing lightcurves in Sec.~\ref{sec:microlensing}.
With the microlensing basics outlined, we describe our method in Sec.~\ref{sec:stats} including: accounting for observation bias (Sec.~\ref{sec:detectProb}), single event classification (Sec.~\ref{sec:singleEventClass}), and fully hierarchical inference (Sec.~\ref{sec:populations}). In Sec.~\ref{sec:verification} we describe our verification testbed which includes a set of simulated population models (Sec.~\ref{sec:pop_model}), a simulated microlensing survey (Sec.~\ref{sec:det_efficiencies}), and performing inference at the single event  (Sec.~\ref{sec:pop_realization}) and population level (Sec.~\ref{sec:forward_modeling}). With our model and simulation framework laid out, we then apply our method to a suite of simulated datasets in Secs.~\ref{sec:source_classification} and \ref{sec:populationInference}, focusing on single event and population-level inferences, respectively. We summarize our findings in Sec.~\ref{sec:conclusions}.

\section{Photometric Microlensing}\label{sec:microlensing}


Consider a point lens with mass $M$, and a more distant point source, at distances $D_{L}$ and $D_{S}$ from an observer, respectively. In the case of perfect lens-source alignment, the gravitational field of the lens deflects the light of the background source forming an Einstein ring of angular radius \citep{Chwolson1924, Einstein1936},
\begin{equation}
\theta_E = \sqrt{ \frac{4 G M }{c^2} \left(\frac{1}{D_L} - \frac{1}{D_S}\right)}\,.
\label{eq:thetaE}
\end{equation}
For imperfect lens-source alignments, two, usually unresolved, images of the source are formed \citep{Liebes1964, Refsdal1964}. As the lens passes between the source and observer the source images change brightness giving rise to an apparent amplification of the background source flux \citep{Paczynski1986},
\begin{equation}
A(t) = \frac{u(t)^2 + 2}{u(t)\sqrt{u(t)^2 + 4}}\,.
\end{equation}
Here, $u(t)$ is the magnitude of the lens-source angular separation vector in units of $\theta_{E}$. The relative lens-source trajectory, $\boldsymbol{u}(t)$, can be parameterized by \citep{Gould2004},
\begin{equation}
    \boldsymbol{u}(t) = \boldsymbol{u_{0}} + \frac{t-t_{0}}{t_{\rm E}}\boldsymbol{
    \hat{\mu}}_{\text{rel}} + \boldsymbol{P}(t; \boldsymbol{\pi}_{\rm E})
    \label{eq:traj}
\end{equation}
Here, $u_{0}=|\boldsymbol{u_{0}}|$ is the magnitude of lens-source impact parameter in units of $\theta_{E}$, $t_{0}$ is the time of lens-source closest approach, $t_{E}=\theta_{E}/\mu_{\rm rel}$ is the Einstein crossing time where $\boldsymbol{\mu}_{\rm rel}$ is the relative lens-source proper motion vector. The first two terms of Eq.~\eqref{eq:traj} make up the standard \cite{Paczynski1986} rectilinear trajectory model. The third term in Eq.~\eqref{eq:traj} accounts for the annual microlensing parallax signal which is caused by the acceleration of an Earth-based observer \citep{Alcock:1995nk}. $\boldsymbol{\pi}_{\rm E}$ is the vector microlensing parallax which can be described by its magnitude $|\boldsymbol{\pi}_{\rm E}|=\pi_{\rm E}=\pi_{\rm rel}/\theta_{E}$, where $\pi_{\rm rel} = 1{\rm au} \; (1/D_{L} - 1/D_{S})$ is the relative lens-source parallax and $\phi$ is the angle between the ecliptic north and the direction of the lens-source relative proper motion in the heliocentric frame.

The expression for $\boldsymbol{P}(t; \boldsymbol{\pi}_{\rm E})$ depends on the on-sky microlensing event location, and in all work that follows we use the results in Section 3.1 of \cite{Golovich:2020acu} which are based on \cite{2013ApJ...763L..35G}. Annual microlensing parallax can lead to typically subtle \citep[e.g.,][]{Alcock:1995nk,Golovich:2020acu, Kaczmarek2022} but sometimes extreme \citep[e.g.,][]{2016MNRAS.458.3012W, Kruszynska2022} asymmetrical deviations from the standard \cite{Paczynski1986} lightcurve.

The flux of the unresolved blended images during the microlensing event can be written as,
\begin{equation}
F(t; \boldsymbol{\theta}) =  F_{\rm Base} + b_{\rm sff}F_{\rm Base}\left[A(t; u_{0}, t_{0}, t_{E}, \pi_{E}, \phi) - 1\right].
\label{eq:flux}
\end{equation}
$F_{\rm Base}$ is the total baseline flux including both the unlensed source flux and all unresolved blended light. $b_{\rm sff}$ is the fraction of unlensed source flux to the total base flux. Overall, we can describe the lightcurve with 7 parameters, $\boldsymbol{\theta}=\{F_{\rm Base},b_{\rm sff}, t_0, t_E, u_0, \pi_{E}, \phi \}$. 

Examination of Eqs. (\ref{eq:thetaE}-\ref{eq:flux}) shows that the only parameters which can be inferred that contain any information on the lens mass, and therefore its identity, are $t_{E}$ and $\pi_{E}$. However, both  $t_{E}$ and $\pi_{E}$ are in units of $\theta_{E}$ which cannot be inferred from the lightcurve in this simple scenario. Overall, this means that there is no direct lens mass information contained in the photometric signal for a single event -- it is degenerate with the relative lens-source distance and velocity.

Prospects for understanding the nature and identity of lenses via photometric microlensing lensing does improve when a large sample of events can be detected over the course of a survey \citep[e.g.,][]{Udalski2015,KMTnet2016,Husseiniova2021}. In this case, different lens types (e.g., Stars, White Dwarfs, Neutron Stars, SOBHs, Free Floating Planets or PBHs) are expected to have differing population characteristics such as different mass distributions, kinematics, and spatial configurations in the Galaxy. These population level differences project down to populations of different lenses producing microlensing events with different characteristics \citep[e.g.,][]{2021arXiv210713697M, 2023arXiv230308280S}.

\begin{figure}
\includegraphics[width=\linewidth]{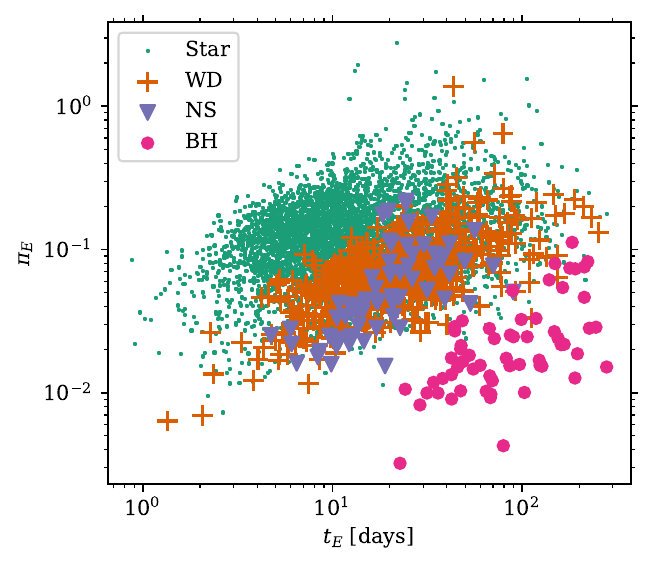}
\caption{
Shown above are the event parameters $t_E$ and $\pi_E$ from a microlensing simulation of the Milky Way bulge produced by {\tt PopSyCLE}, as published by \citet{Lam_2020}.
A strong correlation can be seen between the $t_E$ and $\pi_E$ parameters and mass (class) of the lens.
This correlation will be be key to unraveling the subpopulation makeup of the total population of lensing objects in the galaxy.
}\label{fig:popsycle_te_pie}
\end{figure}

Figure \ref{fig:popsycle_te_pie} shows a simulation of microlensing events in $t_{E}-\pi_{E}$ space from the Population Synthesis for Compact object Lensing Events code \citep[{\tt PopSyCLE};][]{Lam_2020} assuming an OGLE-IV-like microlensing survey. {\tt PopSyCLE} combines galactic and evolutionary models \citep{Kalirai2008, Sharma2011, Sukhbold2016, Raithel2018, Hosek2019} with microlensing survey characteristics to simulate detectable microlensing event catalogs for different lens populations. Figure \ref{fig:popsycle_te_pie} shows that the different lens types do occupy different, albeit overlapping, areas of $t_{E}-\pi_{E}$ space. This separation is fundamentally caused by the scaling of these parameters with respect to the lens mass: $t_E \propto \sqrt{M}$ while $\pi_E \propto 1/\sqrt{M}$. These relationships result in the negative correlation between these two parameters with respect to a changing lens mass, all else being equal. In principle this suggests that given a survey of photometric microlensing events where we can measure $t_{E}$ and $\pi_{E}$ it is possible to make inferences about the different underlying subpopulations of lenses.

\section{Hierarchical Inference with Detection Bias}\label{sec:stats}


To robustly characterize subpopulations of lenses we must account for bias and uncertainty as rigorously as possible - from uncertainty in a single microlensing events' characteristics, to the uncertainty in the identity of the lens for a given event, to having an unknown lens population model. 

We start with the concept of event detection probability in Sec.~\ref{sec:detectProb} and its definition for a single event in the context of a population model. We then move on to 
assigning probabilistic lens classifications to single events in Sec.~\ref{sec:singleEventClass}. Finally, we put everything together in the context of a fully hierarchical population analysis in Sec.~\ref{sec:populations}. The rest of Sec.~\ref{sec:stats} first follows standard results in the literature ~\citep[e.g.,][]{Loredo:2004,Vitale:2020aaz,Mandel:2018mve,Taylor:2018iat} that are then extended to our specific class of models to improve computational tractability.

In what follows, $\eparam$ are the parameters describing a single microlensing lightcurve, defined in Sec.~\ref{sec:microlensing}. $\D$ is the lightcurve data for a single microlensing event - which is a collections of times, fluxes, and flux errors. Generally, $\{\}$ denotes sets, for example, $\{\D\}$ and $\{\eparam\}$ correspond to some set of lightcurves and events parameters, respectively. 
When considering population models, we will parametrize the full model as $\hyp$, representing all parameters relevant to population modeling.
When considering different subpopulations of lenses, we will denote each class by $\classa$, where $a\in [0,\nclass)$ (e.g., stellar, SOBH or PBH lenses).
$N$ is the total number of predicted microlensing events (detected or undetected) by the model. The parameters controlling the subpopulation distributions are $\{\hypshapea\}$, and the parameters controlling the relative abundance of each subpopulation, $\{\popfraca\}$, with $a \in [0,\nclass)$ for $\nclass$ subpopulations.
Examples of $\{\hypshapea\}$ are given in Sec.~\ref{sec:verification}, including parameters of the mass spectrum of lenses. 
However, these parameters can represent any feature of a subpopulation of the population model, not just the mass spectrum.
As $\{\popfraca\}$ are the relative abundances, $\sum_a \popfraca = 1$. 
In summary, $\hyp = N \cup \{\hypshapea\}\cup \{\popfraca\}$. 

\subsection{Detection Probabilities}\label{sec:detectProb}

Detection bias means that our observed set of events are not a fair sample of the true, underlying distribution.
This is because some microlensing events are easier to observe than others.
This effect can be accounted for by defining a ``trigger'', $\trig$, and its probability. %
This means that once event data $\D$ has been recorded at the detector, it either produces a trigger, signifying it is a microlensing event, or not. 
This trigger is evaluated according to whether some deterministic criteria ($\rho(\D)>\rho_{\rm threshold}$) is met that is typically related to a signal-to-noise ratio (SNR) calculation;
\begin{equation}
p(\trig| \D) = 
\begin{cases}
0 & \rho(\D) < \rho_{\rm threshold} \,, \\
1 & \rho(\D) \geq \rho_{\rm threshold} \,. \\
\end{cases}
\label{eq:tigger}
\end{equation}
Here, and in the work that follows we have assumed that the detection criterion is model-independent i.e., $\rho$ only depends on $\D$.

In the context of modeling a population of microlensing events, we can build on our definition of $p(\trig| \D)$ to quantify event detection probabilities. First, we can calculate the probability of a trigger given an event exists with a certain set of parameters $\eparam$. 
As the concept of a trigger is inherently tied to detector noise and detector limitations, we will need to introduce a set of data $\D$ as a parameter and marginalize over all possible noise realizations consistent with the noise model $p(\D|\eparam)$, giving 

\begin{align}\nonumber
p(\trig|\eparam) &= \int p(\trig , \D | \eparam)\;d\D \,,\\
 &= \int  p(\trig | \D ) p(\D|\eparam)\;d\D\,.
 \label{eq:p_trig_theta}
\end{align}

\noindent This detection probability can be computed using Monte Carlo integration, averaging trigger probabilities over all possible data sets consistent with the single event model likelihood, $p(\D | \eparam)$, conditioned on the event properties $\eparam$. 

Using Eq.~\eqref{eq:p_trig_theta}, the detection efficiency for a population model given a set population parameters, commonly defined as $\alpha$ in the literature \citep[e.g.,][]{Mandel:2018mve,Vitale:2020aaz}, is given by,

\begin{align}\label{eq:alpha}\nonumber
p(\trig | \hyp) \equiv \alpha &= \int d \D \int d\eparam p(\trig, \eparam,\D| \hyp) \,,\\
&= \int d \D \int d\eparam p(\trig| \D)p( \D | \eparam) p( \eparam| \hyp)\,.
\end{align}

\noindent This quantity can also be computed via Monte Carlo integration by simulating values of $\eparam$ drawn from the population model and subsequently drawing a noise realization from the event likelihood, which is described in App.\ref{app:alphaCalc}.  Eq.~\eqref{eq:alpha} shows that once the probability of a set of data is conditioned on $\eparam$, the probability of $\D$ is independent of the population model. 
Conceptually, $\alpha$ can be understood as the efficiency of a population model to produce detectable events.
This can also be understood by noting that $\alpha$ can be related to the number of expected detectable events ($N^{\rm det}$) and the total number of expected events ($N$) by the relation $\alpha = N^{\rm det} / N$ \citep{Loredo:2004,Vitale:2020aaz,Mandel:2018mve,Taylor:2018iat}, i.e., that $\alpha$ is the fraction of detectable events over the total number of events.

\subsection{Classification of a single event}\label{sec:singleEventClass}

With the definitions of detection probabilities in hand, we would like to know, given some value of the population parameters $\hyp$,  what is the probability that a single event belongs to a subcategory of the population? 
This leads us to the following posterior probability of an event belonging to a certain class

\begin{equation}\label{eq:classBayes}
p(\classa | \D, \trig, \hyp)= \frac{p(\classa| \hyp)p(\D , \trig | \classa, \hyp)}{p(\D, \trig | \hyp)}\,.
\end{equation}

\noindent This posterior probability is directly related to Bayesian model selection methods. If lens-classification is treated as a model selection problem, the ratio of these posterior probabilities for different classes is the posterior odds. Taking the ratio when neglecting the prior probability of each class yields the Bayes' factor. All of these quantities (the normalized posterior probability, the posterior odds, and the Bayes' factor) are useful metrics at understanding the classification problem, but our focus will be on the normalized posterior probability.

The prior, $p(\classa | \hyp)$, is the probability that a lens belongs to a given class without considering the data. The likelihood, $p(\D , \trig | \classa, \hyp)$, can be simplified. We first re-write the likelihood using the product rule  
\begin{align}\label{eq:selectionEffectsProd}\nonumber
p(\D, \trig | \classa,\hyp) &= p( \trig |\D, \classa,\hyp) p(\D| \classa, \hyp) \,,\\ 
 &=  p(\D| \classa, \hyp) \,,
\end{align}
where we see that the selection effects completely drop out of the equation, i.e., $p(\D,\trig|\classa,\hyp)=p(\D|\classa,\hyp)$. This is because $p(\trig|\D, \classa,\hyp)=p(\trig|\D) =1$ (Eq. \ref{eq:tigger}) for an event that has been detected. 

Going from the first line in Eq.~\eqref{eq:selectionEffectsProd} to the second, it is tempting to write $p(\trig | \D, \classa, \hyp) $ as something related to the survey efficiency functions commonly published along with survey data.
However, this is incorrect, as this probability is also conditioned on the data $\D$, which takes precedence. 
While some microlensing surveys select events based on model parameters like $u_0$ and $t_E$ \citep[e.g.,][]{Husseiniova2021}, it should be noted that these are maximum likelihood estimates completely based on the characteristics of the data.
These fitted parameters do not share the same meaning as the parameters $\eparam$, which are implicit when conditioning on $\classa$ in Eq.~\eqref{eq:selectionEffectsProd}.
Instead, we note that the correct interpretation is $p(\trig|\D)=1$ for an event that is known to have a trigger, regardless of event model parameters or event classification.
This means that the detection efficiency \emph{does not} play a role in the likelihood of individual events\footnote{While the fact that selection effects should be neglected when analyzing single events is a known result \citep[e.g.,][]{Mandel:2018mve}, showing its derivation is further validation of our methodology before moving on to hierarchical analyses, where selection effects do enter. Furthermore, our derivation illustrates how selection effects do not enter when considering the context of classification, an analysis which utilizes hierarchical information but does not infer hierarchical information, which is less obvious.}.

Assuming that our set of considered lens subpopulations is complete, the evidence of a single lens (the denominator of Eq.~\eqref{eq:classBayes}) is,

\begin{equation}
p(\D, \trig | \hyp) = \sum_b p(\classb|\hyp) p(\D|\classb, \hyp)\,,
\end{equation}

\noindent where we are summing over the finite and complete set of lens classes. Here we have also used again the fact that $\trig$ only depends on $\D$, and therefore $p(\D,\trig|\classa,\hyp) = p(\D|\classa,\hyp)$, from the arguments above.
This leads to the following identification: $p(\D, \trig|\hyp) = p(\D|\hyp)$, or that the presence of a trigger does not carry any additional information not already contained in the stream of data itself.

After simplifying the terms in Eq. (\ref{eq:classBayes}), the dependence on $\trig$ disappears, so that $p(\classa | \D,\trig, \hyp) = p(\classa|\D,\hyp)$. Fundamentally, this is because selection effects are the embodiment of factors that lead to some signals truly in the dataset being classified as an event ($p(\trig|\D)=1$) and others to be missed ($p(\trig|\D)=0$). 
When considering one event (already designated as a detection), selection effects play no role even when performing the analysis in the context of population models.
However, as we will see in the population analysis in Sec.~\ref{sec:populations}, $\trig$ enters into the formalism when accounting for the fact that the full dataset being considered is incomplete.

We can now write Eq.~\eqref{eq:classBayes} in a form that can be computed by introducing $\eparam$,

\begin{align}\label{eq:finalPosteriorclass}\nonumber
p(\classa | \D, \hyp) &= \frac{p(\classa| \hyp)}{p(\D| \hyp)} \\
&\times \int p(\D | \eparam ) p(\eparam |\classa, \hyp)d\eparam \,.
\end{align}

Practically, we can compute the integral on the right hand side by importance sampling if we have $S$ independent posterior samples $\eparam_{c}\sim p(\eparam|\D)$ drawn under some prior, $\pi(\eparam)$, with wide support \citep{2010ApJ...725.2166H},
\begin{align}\label{eq:finalPosteriorclassIS}\nonumber
\int p(\D | \eparam ) &p(\eparam |\classa, \hyp)d\eparam \approx  \\
&\frac{1}{S} \sum_{c=0}^{S} \frac{ p(\eparamc |\classa, \hyp)}{\pi(\eparamc)}\,.
\end{align}
Here, the evidence for the single event analysis $p(\D)$ was absorbed into the updated evidence $p(\D| \hyp)$ as an overall constant.
With Eq.~\eqref{eq:finalPosteriorclassIS}, we can leverage previously calculated posterior samples to address the question of lens classification for a single event.

\subsection{Population Analysis}\label{sec:populations}

We now turn to inferring $\hyp$ using a set of $\NOBS$ different microlensing events, $\{\DDi\}$, and detection information $\{\trig\}$. The posterior probability density of $\hyp$ is well documented in the literature ~\citep[e.g.,][]{Loredo:2004,Vitale:2020aaz,Mandel:2018mve,Taylor:2018iat}, so we state the result here and leave a detailed derivation to App.~\ref{app:hierarchicalDeriv}. We have,

\begin{subequations}\label{eq:popPost}
\begin{equation}
p(\hyp | \{\DDi\}, \{\trig\},\NOBS) \propto \frac{p(\hyp)e^{-\alpha N}N^{\NOBS}  }{p(\{\DDi\}, \{\trig\}, \NOBS)}\prod_{i=0}^{\NOBS}\mathcal{L}_i^{{\rm obs}}\,,
\end{equation}
\begin{equation}
\mathcal{L}_i^{{\rm obs}}  = \int d\eparamdi p(\DDi|\eparamdi)p(\eparamdi|\hyp)\,.
\end{equation}
\end{subequations}

Here, $p(\hyp)$ is the population parameter prior and $p(\{\DDi\}, \{\trig\}, \NOBS)$ is the evidence. The factor $e^{-\alpha N}N^{\NOBS}$ follows from assuming events are generated via an (inhomogeneous) Poisson process ~\citep[e.g.,][]{2011ApJ...742...38Y,Loredo:2004}, which penalizes population models that do not predict the correct number of detected events ($\alpha N$). This factor accounts for selection bias by marginalizing over the unknown events in the data that fail to rise above the detection threshold. 
The quantity $\mathcal{L}_i^{\rm obs}$ is the marginalized event likelihood, that is independent of selection effects for the reasons outlined in Sec.~\ref{sec:singleEventClass}.

The expression derived in Eq.~\eqref{eq:popPost} differs from those of past works, for example \citet{2021arXiv210713697M} (see their Eq.~2) and \citet{2023arXiv230308280S} (see their Eqs.~4 and~10, where the total likelihood is the product of these two expressions). 
Both of these works neglect information about the overall rate of events (including the effects of Poisson statistics), which can be an important piece of information when disentangling population information.
Eq.~\eqref{eq:popPost} generalizes these past methodologies such that differential rate information, which can differ dramatically between subpopulations of lenses, is formally included in the analysis.
In the case that rate information is still deemed unnecessary, it can be marginalized out of Eq.~\eqref{eq:popPost} formally, with an appropriate choice of prior \citep{Fishbach:2018edt}.
Furthermore, we generalize the work of \citet{2023arXiv230308280S} by deriving a more nuanced treatment of selection affects, replacing the detection efficiency factor of their work with the integrated detection efficiency $\alpha$, defined in Eq.~\eqref{eq:popPost}.
\citet{Mandel:2018mve} showed that the treatment of \citet{2023arXiv230308280S} can lead to biased conclusions when considering systems with strong selection effects.
While these past works were less susceptible to these differences because of a restricted focus on certain ranges of timescales, both of these extensions become increasingly important when considering a fully global analysis of the data, simultaneously considering all subpopulations with overlapping predictions of event parameters.

In addition to the full model in Eq.~\eqref{eq:popPost}, we also explored the use of a restricted model. In this restricted model, we exploit the fact that we are using a mixture model for lens classes and fix the parameters of the lens subpopulations $\{\hypshapea\}$, only allowing the subpopulation mixing fractions $\{\popfraca\}$ to vary. While this method is $\approx10^{2}$ times faster to compute than the full model, it can be susceptible to biased inferences for our problem. We detail the restricted model's exploration in App. \ref{app:restricted} as it might be of use to other astrophysical problems or for future microlensing applications once these subpopulations are better understood.

With Eq.~\eqref{eq:finalPosteriorclass} and Eq.~\eqref{eq:popPost}, we have the machinery to understand individual events in the context of their populations and extract hierarchical information from noisy, biased surveys robustly. 


\section{Verification Design}\label{sec:verification}


In this section we describe the process of validating our proposed methods. To do this, we simulate microlensing events from our population models along with a microlensing survey. We then attempt to recover and disentangle the injected lens populations with our method. This self-consistent testbed enables the evaluation of our methods efficacy in an environment free of systematic bias.

As a specific test case, we consider five different population models denoted by $\{\hyp_0,\hyp_1,\hyp_2,\hyp_3,\hyp_4\}$, constructed to mimic features of the BH mass spectrum being reported through gravitational wave observation by the LIGO-Vigo Collaboration (LVC) collaboration~\citep{LIGOScientific:2020kqk,LIGOScientific:2021psn}. Namely, we utilize a BH mass spectrum with a power law component and a Gaussian component, as a model with these features yielded the best fit from LVC's analysis. With data generated from these population models, we use our methodology to determine if these features could be consistent with a subpopulation of PBHs and detected via microlensing. As a byproduct of these tests, we evaluate the ability of this method to constrain aspects of the stellar and SOBH subpopulations. Below, we outline the population models used in Sec.~\ref{sec:pop_model}, and our simulated microlensing survey in  Sec.~\ref{sec:det_efficiencies}. Finally, we describe the numerical methods used to construct and analyze a population realization in Sec.~\ref{sec:pop_realization} and Sec.~\ref{sec:forward_modeling}, respectively.

\subsection{Population Models}\label{sec:pop_model}

We consider three intrinsic lens subpopulations (i.e., before observation) - Stars, SOBHs and PBHs, which only differ in their mass spectra. It is critical to note, however, that this framework can be extended to include inference on all hyperparameters of the lens subpopulations, including lens/source velocity distributions, spatial distributions, etc. Marginalizing over these uncertainties will be crucial to applying these methods to real data, but that extension is left to future work. Starting with the mass distribution adequately shows the effectiveness of this approach. 

We structure every population model in the same way, only varying the number of PBH sources in the data. Common to all population models considered in the verification process, the stellar and SOBH subpopulations are represented by a Pareto type-II power law mass distribution implemented in SciPy \citep{2020SciPy-NMeth} starting at $0.07M_{\odot}$ and $5M_{\odot}$, respectively. These mass distributions have the form,

\begin{equation}\label{eq:pareto}
p(M | b, \loc, \scale) = 
\begin{cases}
\frac{ b}{ \scale} \left(\frac{ \left(M-\loc\right)}{\scale}\right) ^{-b - 1},\, &M> \mu+\sigma\,, \\
0,\, &M\leq \mu+\sigma\,, \\
\end{cases}
\end{equation}

\begin{figure}
\includegraphics[width=\linewidth]{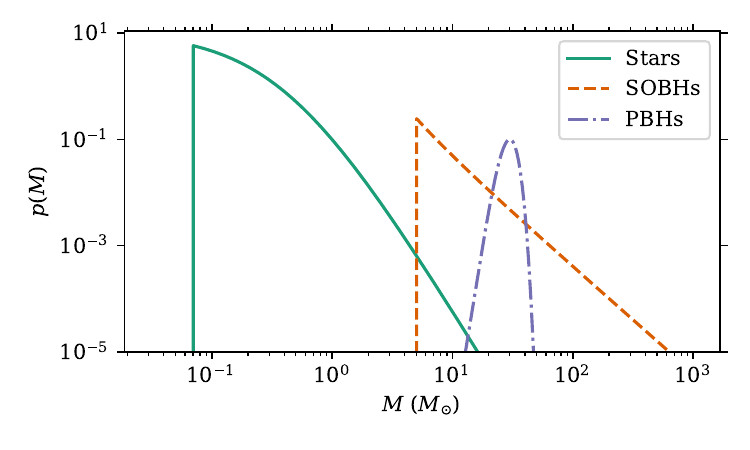}
\caption{
	The true mass distributions used in our toy model universe.
	The distribution is comprised of three subpopulations, meant to reflect realistic distributions in nature.
    Note, the subpopulation mass distributions are normalized independently, so the relative amplitudes are not indicative of what was used to produce the data.
}\label{fig:analytic_mass_pdf}
\end{figure}

Here, $M$, $\loc$ and $\scale$ are in solar masses. $M_{\rm min} = \loc + \scale$ is the minimum mass. 
The parameter $b$ is the tail parameter for a Pareto distribution, and is related to the spectral index of a power law through the relation $b+1$.
The PBH subpopulation is described by a Gaussian centered at $30M_{\odot}$ with mean ($\mu$) and standard deviation ($\sigma$) parameters. The adopted parameter values and shapes of these mass spectra are shown in Table~\ref{tab:true_population} and Fig.~\ref{fig:analytic_mass_pdf}, respectively. For numerical stability we also require all lenses to have $M<1000 M_{\odot}$ which is well over the pair instability mass limit of SOBHs~\citep{VINK:2015,2002ApJ...567..532H,2003ApJ...591..288H}. 

In addition to lens masses, to generate microlensing events in all models we assume: $D_{L}$, $D_{S}$ $\sim \mathcal{U}(2000{\rm pc}, 8000{\rm pc})$ where $D_{L} < D_S$, $b_{\rm sff} \sim \mathcal{U}(0,1)$, and $\phi \sim \mathcal{U}(0,2\pi)$, which were chosen to be reasonably physical, simple distributions. We assume events have a baseline magnitude $I_{\rm Base} \sim \mathcal{N}(\mu = 21.066 , \sigma = 1.780)$ with a reference point of $F_{\rm ref}=1$ corresponding to $I_{\rm ref}=22$ and $\log_{10} \mu_{\rm rel} \sim \mathcal{N}(\mu = 0.81 , \sigma = 0.21)$, which were chosen to be consistent with the {\tt PopSyCLE} simulations \citep{Lam_2020}. Finally, we assume events happen along a random line of sight, giving a variety of parallax orientations, and random peak magnification time $t_0$ uniformly distributed from $0$ to $3650$ days. 
For all models, we fix the number of detected stellar and SOBH lenses to $N^{\rm det}_{\popA}=3225$ and $N^{\rm det}_{\popB}=27$, respectively, to be reasonable approximations for what to expect from current microlensing surveys like OGLE while remaining computationally tractable. 
Averaged over many realizations, this corresponds to $N_{\popA} =15530$ and $N_{\popB}=100$ (detected or undetected) events given each subpopulation's population efficiency, $\alpha_{\popA}$ and $\alpha_{\popB}$ (also an average quantity), and noting the relationship between the two: $N_{a} =N^{\rm det}_{a} / \alpha_a $.

The only difference between all the population models being described in this work is the relative contribution of the PBH Gaussian bump in the mass spectrum. For $\hyp_0$, PBHs contribute roughly $100\%$ of the dark matter in our galaxy ($f_{\rm PBH}=1$) with a PBH relative abundance of $\fracC = 0.032$ \citep{Pruett2022}. 
While this size of a PBH subpopulation is ruled out by observation and experiment for this range of masses, it provides a good point of reference for the analyses of the other population models.
For $\hyp_{1-4}$, we progressively step down the contribution of the PBH subpopulation until $f_{\rm PBH} = 0$ (see Table~\ref{tab:detectionNumbers}).

\begin{table}
{
\centering
\begin{tabular}{ P{2.5cm} | P{2.5cm}}
\hline\hline 
Shape Parameter & Value \\ \hline \hline
$b_{\popA}$ &$2.717$ \\ 
$\loc_{\popA}$ & $-0.398 M_{\odot}$ \\ 
$\scale_{\popA}$ & $0.468 M_{\odot}$ \\\hline
$b_{\popB}$ & 1 \\ 
$\loc_{\popB}$ & $1 M_{\odot}$ \\ 
$\scale_{\popB}$ & $4 M_{\odot}$ \\\hline
$\loc_{\popC}$ & $30 M_{\odot} $\\ 
$\scale_{\popC}$ & $4 M_{\odot}$\\
\hline\hline 
\end{tabular}
}
\caption{
	Above are shown the true model parameters used by each subpopulation (stellar, SOBH and PBH), where the stellar and SOBH subpopulations are modeled as power law distributions and the PBH subpopulation is modeled as a Gaussian distribution.
	The values were picked to reflect realistic scenarios, either inspired by PopSyCLE simulation or from the literature.
}\label{tab:true_population}
\end{table}

\begin{table}
\begin{center}
\begin{tabular}{ P{1.5cm} |P{1.0cm} |P{1.0cm} |P{1.0cm} |P{1.0cm} }
\hline\hline 
Population Model & $N_{\rm Star}^{\det}$ ($\fracA$) & $N_{\rm SOBH}^{\det}$ ($\fracB$) & $N_{\rm PBH}^{\det}$ ($\fracC$) & $f_{\rm PBH}$\\ \hline\hline
$\hyp_0$ & $3225$ ($0.96$) & $27$ ($0.006$) & $143$ ($0.032$) & $1$ \\\hline
$\hyp_1$ & $3225$ ($0.98$) & $27$ ($0.006$) & $41$ ($0.0096$) & $0.29$\\ \hline
$\hyp_2$ & $3225$ ($0.99$) & $27$ ($0.006$) & $12$ ($0.0028 $) & $0.08$\\\hline
$\hyp_3$ & $3225$ ($0.98$) & $27$ ($0.006$) & $3$ ($0.007$) & $0.02$\\\hline
$\hyp_4$ & $3225$ ($0.99$) & $27$ ($0.006$) & $0$ ($0$) & $0$\\
\hline\hline 
\end{tabular}
\end{center}
\caption{
	The table above shows the number of detected sources in the final catalog of each population model used in this study, along with the corresponding relative abundance. 
    The final column shows the correpsonding DM fraction $f_{\rm PBH}$ given the assumptions of \citet{Pruett2022}.
	We denote each population model as $\Lambda_i$ for the $i$-th model, where the only difference in the model used to generate the data was the number of PBH sources.
	Note that the relative abundances shown here are for the intrinsic population model, not the fraction of \emph{detected} sources.
}\label{tab:detectionNumbers}
\end{table}

\begin{figure}
\includegraphics[width=\linewidth]{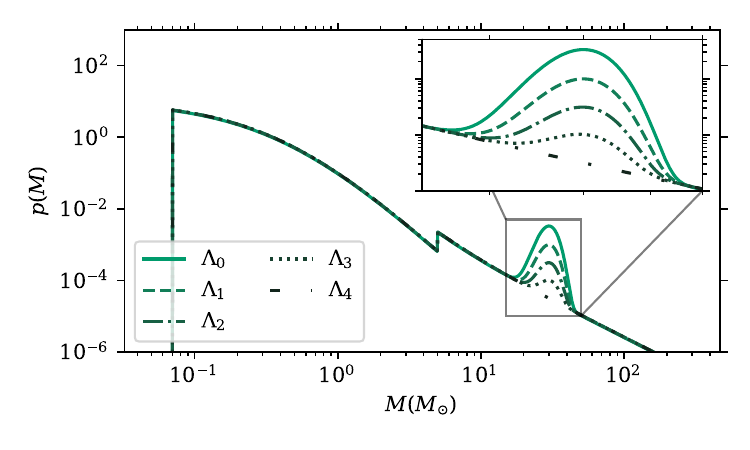}
\caption{
	The mass probability densities for each of the subpopulation models described in Table~\ref{tab:detectionNumbers}.
	Each set of data $\hyp_i$ was created with the same models for the stellar and SOBH subpopulations, while the PBH subpopulation was logarithmically decreased from $f_{\rm PBH}= 1$ to $f_{\rm PBH} = 0$.
}\label{fig:ridgeplot}
\end{figure}

\subsection{Survey Design and Selection Criteria}\label{sec:det_efficiencies}

For our toy microlensing survey, we adopt OGLE-like \citep{Udalski2015} characteristics; a ten year survey with a 3 day cadence and a magnitude measurement error of $\sigma_N = 0.1$ mag for all magnitudes. We neglect gaps in the data from seasonal observations, leaving more realistic survey designs to future work. 
While this cadence is more suggestive of OGLE-III than OGLE-IV and might affect the short timescale end of the $t_E$ distribution, this work focuses on the long timescale end of the distribution. 
Small changes to the observation cadence in the range of hours to days should not drastically impact the long timescale end of the $t_E$ distribution, as evidenced by a comparison of the OGLE-IV detection efficiency curve and the detection efficiency curve in these simulations shown in Fig.~\ref{fig:pdet_curve} above ${\sim} 10$ days.
Furthermore, the choice of using $\sigma_N = 0.1$ mag is \emph{conservative} when considering the estimated variance of the magnitude measurements from \citet{Mr_z_2019} (Fig.~9). While remaining in the white, Gaussian, and stationary noise limit, using a more accurate estimate of the magnitude measurement variance which varies with magnitude will only improve the conclusions of this work.
For detection thresholds, we use a simplified version of the OGLE-IV~\citep{Mr_z_2019} criteria. Namely, for each lightcurve, we take the maximum flux, $F_{\rm max}$, and calculate the average baseline flux, $F_{\rm base}$, and variance of the baseline, $\sigma_{\rm base}^2$, using the data greater than 360 days away from $F_{\rm max}$ (cutting out a 720 day window). We calculate the significance through the $\chi_{3+}$ parameter, defined as 
\begin{equation}\label{eq:bumpSig}
\chi_{3+} = \sum_i \frac{(F_i - F_{\rm base})}{\sigma_{\rm base}}\,,
\end{equation}
where $i$ indexes the flux measurements and begins at $F_{\rm max}$, including all consecutive points above $F_{\rm base}+3\sigma_{\rm base}$.
If $\chi_{3+}<32$, we classified the event as a non-detection. For an event to be detected we also require it to have baseline magnitude less than $21$, and the corresponding change in magnitude between $F_{\rm max}$ and $F_{\rm base}$ to be $>0.1$ mag.

\subsection{ Population Model Realization and Single Event Inference}\label{sec:pop_realization}

\begin{figure*}
\includegraphics[width=\linewidth]{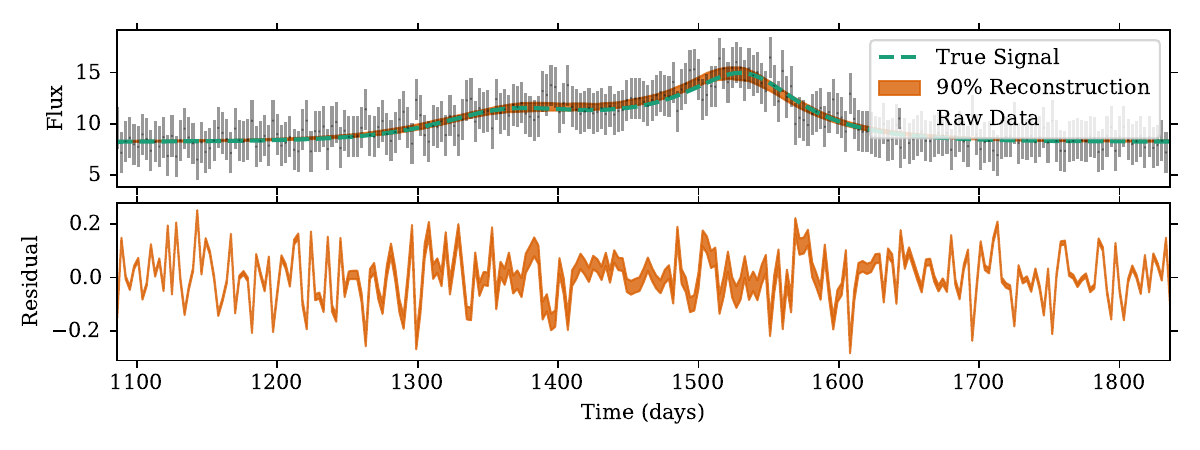}
\caption{
	The data and reconstruction from our analysis for a specific event in the synthetic catalog.
	The top panel shows the raw data (in gray), the true signal in the data (dashed green) and the $90\%$ confidence reconstruction of the signal from our posterior distribution (shaded orange).
	In the bottom panel, we show the residual, defined by the difference between the reconstruction and the data, divided by the average of the reconstruction and the data.
}\label{fig:lightcurveReconstruction}
\end{figure*}

To create a simulated catalog of microlensing events we draw samples of $\eparam$ from the population model. For each of these microlensing events parameters we simulate a light curve according to Sec. \ref{sec:det_efficiencies}, corrupting the data with white Gaussian noise with standard deviation of $\sigma_{N}=0.1$ mag. If the events meet the detection criteria in Sec.~\ref{sec:det_efficiencies}, we add it to our microlensing event catalog. This process is continued until we have $N^{\rm det}_a$ events for each subpopulation, $a \in \{\popA,\popB, \popC\}$ (as outlined in Table.~\ref{tab:detectionNumbers}). 
This is regardless of what $N_a$ should be for each subpopulation as calculated by $N^{\rm det}_a/\alpha_a = N_a$, which is the average quantity over many realizations.
By fixing $N^{\rm det}_a$ for each subpopulation instead of the intrinsic number $N_a$, we are able to more directly compare the outputs of the various realizations.
Fig.~\ref{fig:lightcurveReconstruction} shows an example of a simulated event that meets detection criteria with $t_E = 125$ days and $\pi_E = 0.35$.

To obtain posteriors samples of $\eparam$ for the events in our simulated catalog we first transform certain parameters to log space ($\{\log F_{\rm Base}, b_{\rm sff}, \log t_0, \log t_E, \log u_0, \phi, \log \pi_E\}$) to increase sampling efficiency. We then use a custom Markov chain Monte Carlo (MCMC) sampler defined in ~\citep{Perkins:2021mhb,Perkins:2022fhr}, which has been validated in $15+$ dimensional parameter spaces with jagged, multi-modal features in the posterior. This sampler is built on the concept of parallel tempering ~\citep{PhysRevLett.57.2607,B509983H} to efficiently explore multi-modal posteriors, and utilizes Fisher information matrices to construct efficient proposal densities. In each run of the sampler, random draws from the prior are used at starting points to ensure we are not biased by starting at the known true values and the sampler is run until $\approx1000$ independent samples are collected which is determined by the chain auto-correlation length.

\begin{figure*}
\includegraphics[width=.48\linewidth]{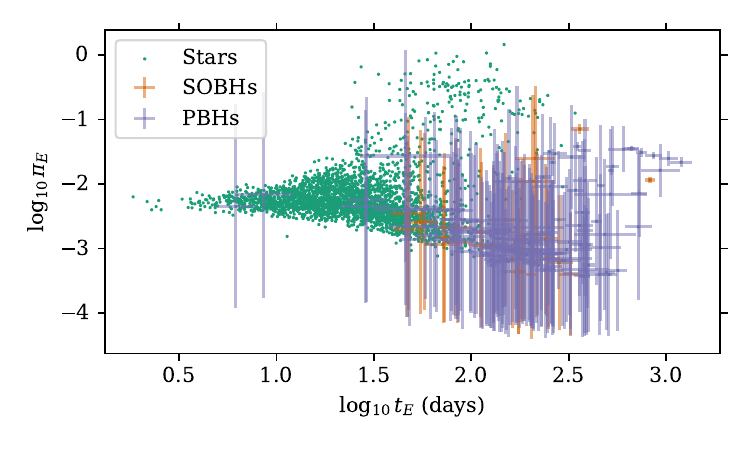}
\includegraphics[width=.48\linewidth]{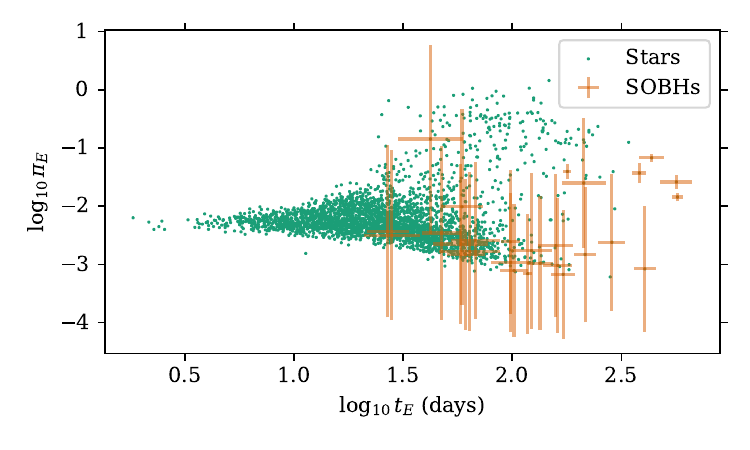}
\caption{
	The median values of the posteriors for all the events in the synthetic catalog produced by population model $\hyp_0$ (left) and $\hyp_4$ (right) are scattered above, separated out by the subpopulation.
	For the SOBH and PBH subpopulations, we show the error bars ($1\sigma$), as calculated by our posteriors from the single event analysis.
    As a reminder, $\hyp_4$ does not include PBH lenses at all. 
	Note the separation of heavier lenses (SOBH and PBH) down and to the right in $t_E$-$\pi_E$ space, as expected.
	Roughly speaking, the distribution of events in this lower right quadrant of this space would give insight into the BH subpopulations of lensing objects in our galaxy.
}\label{fig:tEPiEDistributionErrorBars0}
\end{figure*}

For our prior distributions, we used priors we deemed to be appropriately uninformative for each parameter.
This meant sampling uniformly in the flux $F_{\rm Base}$, the time of maximum magnification $t_0$, the impact parameter $u_0$\footnote{We restrict our analysis to $u_0>0$ which neglects degeneracies leading to multi-modal posteriors \citep{Gould2004}. We can do this in the current study as the simulated data was also restricted to $u_0>0$, ensuring no bias. Analysis of real data will have to be more careful to asses the effects of multi-modality \citep[e.g.,][]{Kaczmarek2022}. Although, its worth noting that work focused on compact objects is less sensitive to these issues as compact object preferentially fall into the low $\pi_E$ space, softening the degeneracy and leading to the joining of these different modes.}, the blending fraction $b_{\rm sff}$, and angle $\phi$.
We sampled uniformly in the \emph{logarithm} of $t_E$ and $\pi_E$ (i.e., uniform in \emph{scale} for these parameters) to ensure proper exploration across the entire parameter space.
The priors are summarized in Table~\ref{tab:priors}.
We assume a likelihood consistent with our simulated Gaussian white noise model\footnote{This ensures no systematic bias for this preliminary study. Of course, future work could begin to relax this requirement to study systematics or more complicated noise models expected in actual data \citep{Golovich:2020acu}},

\begin{equation}
\ln \mathcal{L}(\D|\eparam) \propto -\frac{1}{2} \sum_t^{L_{\rm data}} \frac{(\Dt- I(\eparam,t) )^2 }{ \sigma_n^2}\,,
\end{equation}

\noindent where the index $t$ runs across the entire data $L_{\rm data}$ and $I(\eparam, t)$ is the prediction for the magnitude from our microlensing model, as a function of the event parameters and time. 
Fig.~\ref{fig:lightcurveReconstruction} shows the reconstruction of an example lightcurve in the synthetic catalog with the reconstruction from this analysis overlaid.
Fig.~\ref{fig:tEPiEDistributionErrorBars0} shows the inferred posterior in $t_E-\pi_E$ space for the catalog of events drawn from $\hyp_0$ and $\hyp_{4}$. 

\begin{table}
{
\begin{center}
\begin{tabular} { P{1.5cm}| P{2.5cm} | P{3.5cm} }
\hline\hline 
Parameter & Prior form & Range\\\hline\hline
$F_{\rm Base}$  & Uniform & $[10,5000]$ arbitrary units \\
$b_{\rm sff}$  & Uniform & $[0,1]$ \\
$t_0$  & Uniform & $[0,3650]$ days \\
$t_E$  & Log-Uniform & $[0.5,3000]$ days \\
$u_0$  & Uniform & $[10^{-5},3]$ \\
$\phi$  & Uniform & $[0,2 \pi]$ \\
$\pi_E$  & Log-Uniform & $[10^{-5},3]$ \\
\hline\hline 
\end{tabular}
\end{center}
}
\caption{
	The prior distributions utilized in this study for the individual event model parameters.
	The ranges are listed in the same units as the parameters in the first column (e.g., we sample in $\log t_E$ and the prior is log-uniform for $t_E$, but the range is written here in days).
}\label{tab:priors}
\end{table}

\subsection{Population Inference}\label{sec:forward_modeling}

For the population-level parameters, we use the same MCMC software outlined in Sec.~\ref{sec:pop_realization} to obtain posterior samples. 
To perform the inference, we will need to forward model a population model. In this work, we will be using the same simple models outlined in Sec.~\ref{sec:pop_model}, but when using real data, population simulations like \texttt{PopSyCLE} will need to be implemented (although this pipeline is independent of the exact forward model method being employed). 
We perform inference with two separate population models: once using all three distributions (producing an unbiased estimate free of systematics) and once using only just Stars and SOBHs. The $\hyp_4$ catalog is the only set of simulated data that can be perfectly modeled by only stellar and SOBH subpopulations, as it contains no PBHs. In the case of the other simulated datasets, $\hyp_{0-3}$, modeling with just stellar and SOBH subpopulations introduces systematic bias enabling us to understand how a PBH subpopulations signal could be detected (or evade detection). Overall, $\hyp = \{b_\popA,\loc_\popA,\scale_\popA, N_\popA,$ $b_\popB,\loc_\popB,\scale_\popB, N_\popB\}$ and $\hyp = \{b_\popA,\loc_\popA,\scale_\popA, N_\popA,$ $b_\popB,\loc_\popB,\scale_\popB, N_\popB,$ $\loc_\popC,\scale_\popC, N_\popC\}$ for the model containing just Stars and SOBHs and for the model containing all lens subpopulation, respectively.

The priors used for $\hyp$ are detailed in Table ~\ref{tab:priorBounds} and were chosen to be uninformative, uniform, and with boundary values commensurate with the uncertainty in the prior understanding of the subpopulation of stars and SOBHs. Additionally, for the SOBH subpopulation, we stipulate that the minimum mass of the power law distribution must be in the range of $[2,6]M_{\odot}$, as there is observational evidence for this upper limit from GW observations, X-ray binaries, and radial/photometric observations~\citep{KAGRA:2021duu,Mapelli:2018uds,2010ApJ...725.1918O,2019Sci...366..637T}. 

\begin{table}
{
\begin{center}
\begin{tabular}{ P{1.5cm} | P{1.5cm} | P{1.5cm} | P{1.75cm} }
\hline\hline 
Parameter &  Unit &Low Limit & High Limit \\\hline \hline
$b_\popA$ & - & 0.1 & 10 \\
$\loc_\popA$ & $M_{\sun}$ & -5 & 5 \\
$\scale_\popA$ & $M_{\sun}$ & 0 & 5 \\
$N_\popA$ & - &0 & $2 N_{\rm data}$\\
$M_{{\rm min},\popA}$ & $M_{\sun}$ & 0.01 & 1 \\\hline
$b_\popB$ & - & 0.1 & 10 \\
$\loc_\popB$ & $M_{\sun}$ & -20 & 20 \\
$\scale_\popB$ & $M_{\sun}$ & 0 & 20 \\
$N_\popB$ & - & 0 & $2N_{\rm data}$ \\
$M_{{\rm min},\popB}$ & $M_{\sun}$ & 2 & 6 \\\hline
$\loc_\popB$ & $M_{\sun}$ & 1 & 80 \\
$\scale_\popC$ &  $M_{\sun}$ & 1 & 20 \\
$N_\popC$ & - & 0 & $2N_{\rm data}$ \\
\hline\hline 
\end{tabular}
\end{center}
}
\caption{
	The prior ranges used in the inference of the simulated population data.
	All priors are uniform, with boundaries set by the values in this table.
	The quantity $N_{\rm data}$ represents the total number of events (detected or not) predicted by the true population model.
	This number ranges between $15630$ and $16160$.
}\label{tab:priorBounds}
\end{table}


\section{Application to Single Events}\label{sec:source_classification}


We first explore the implications of treating the class of single events probabilistically and compare it to example, typical cuts in $\pi_{E}-t_{E}$ \citep[e.g.,][]{Golovich:2020acu} to classify events. Specifically, for the purpose of identifying BHs, we compare our method with a linear cut in $\log_{10}t_E$-$\log_{10}\pi_E$ space defined such that $50\%$ of the events with posterior medians below the line are classified as BHs according to simulations from a population model, which is used by \citep{Golovich:2020acu}. 

\begin{figure}
\includegraphics[width=\linewidth]{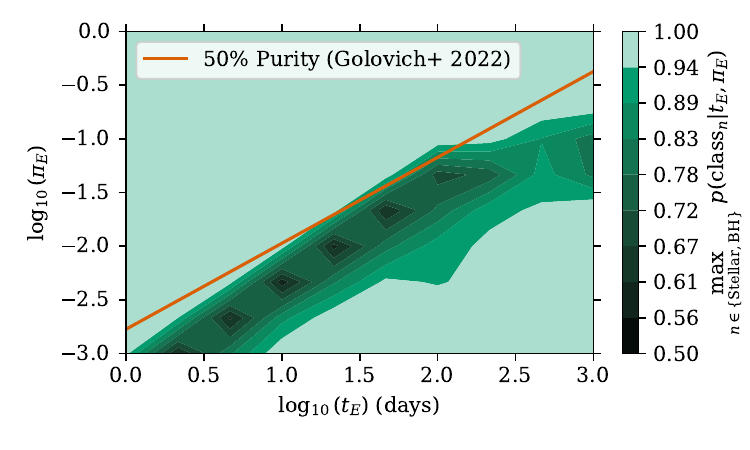}
\caption{
    Shown above are contours illustrating the regions of intrinsic uncertainty in the classification problem compared to the $50\%$ purity as calculated consistent with \citet{Golovich:2020acu}.
    The contours are derived by taking the maximum probability between an event to include either a stellar or BH (both SOBH and PBH) lens. 
    These are calculated assuming \emph{perfect} measurement, i.e., that the posterior for the event parameters $t_E$ and $\pi_E$ are delta functions.
    The regions of dark shading illustrate regions in parameter space where intrinsic overlap in the predictions from different subpopulations fundamentally limits our ability to classify an event with these methods. 
    With photometry alone, classifying events that fall in the dark green regions of parameter space will not be improved drastically even with infinite observational precision.
    The light regions reflect parts of parameter space where classification is highly certain.
    We note that the exact location of the contours fluctuate with numerical noise in our simulation, but the general structure is robust across different realizations.
    This is particularly true in regions with low expected rates, like high $\pi_E$ and high $t_E$.
}\label{fig:probabilityOfIdentification}
\end{figure}

Fig.~\ref{fig:probabilityOfIdentification} shows a comparison of these two lens classification methods in the $\pi_{E}-t_{E}$ space for $\hyp_{4}$. Specifically, the $50$\% purity line is overlaid on the contours calculated by maximizing the class probability across the stellar and BH lenses (both SOBH and PBH). 
Fig.~\ref{fig:probabilityOfIdentification} reveals that our method captures high-order structure in the intrinsic uncertainty in the lens class predictions from the population models missed with the $50\%$ purity line. 

Firstly, Fig.~\ref{fig:probabilityOfIdentification} shows that there are regions of $\pi_{E}-t_{E}$ space that do not trace the 50\% purity line. Intrinsic to the population model itself, an event cannot have more than $\sim 50\%$ lens class confidence of a BH vs a Star (dark regions), even if $\pi_{E}$ and $t_{E}$ are known perfectly. This has implications for allocating followup resources for events in progress, because if an event lies in one of the dark regions, then taking further high-cadence and high-precision photometric data as the event the event evolves to shrink the $\pi_{E}-t_{E}$ posterior will not improve lens-class confidence. Conversely, events with diffuse constraints on $\pi_{E}-t_{E}$ that are in areas of light contours could have their class confidence improved with followup observations.

\begin{figure}
\includegraphics[width=\linewidth]{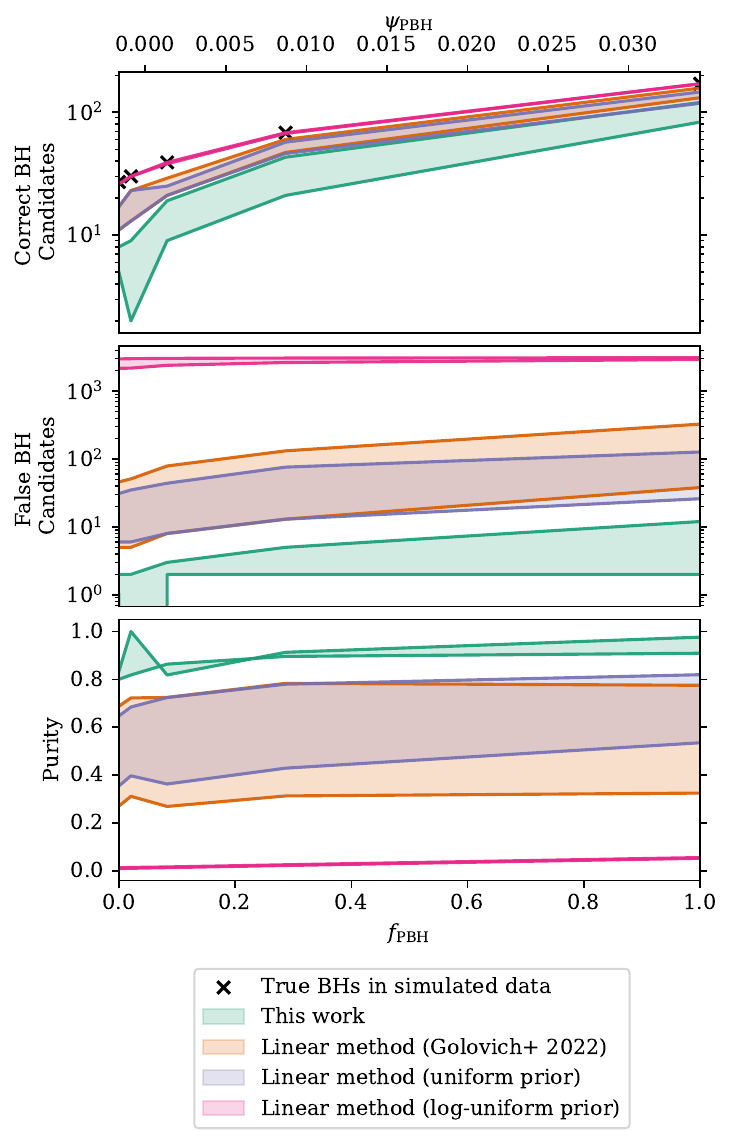}
\caption{
	\textbf{Top}: the number of BH lenses correctly identified as BH candidates for each simulation according to each selection method (true positives): the method of this paper (green), the linear method using the priors from \citet{Golovich:2020acu} (orange), the linear method using linear priors in $t_E$ and $\pi_E$ (purple), and the linear method using log-uniform priors in $t_E$ and $\pi_E$ (pink).
    The width of the band is calculated using two different tuning criteria.
    In the case of the method proposed in this work, we used a threshold probability of $p(\classBH|\D,\hyp)>0.9$ and $p(\classBH|\D,\hyp)>0.5$.
    In the case of the linear method, we used lines designed to have $50\%$ and $90\%$ purity.
    An important note is that this figure should not be interpreted as a method for measuring the abundance of BHs. 
    The abundance is a population-level parameter and is more appropriately handled in the analysis of Sec.~\ref{sec:populationInference}.
    \textbf{Middle}: the number of stellar lenses incorrectly identified as BH candidates for each simulation for each selection method (false positives).
	\textbf{Bottom}: the purity of these classifications, defined as the number of correctly identified BH candidates divided by the total number of BH candidate classifications.
    We see the probabilistic classification method of this paper outperforms the linear method with any of the single-event priors considered here, in terms of the highest purity.
    Furthermore, the probabilistic method is fundamentally independent of the priors used in the single event analysis and is generally robust to changes in the arbitrary threshold parameter (in terms of purity).
}\label{fig:ratePurity}
\end{figure}

To quantify the advantage of using our probabilistic lens classifications over purity cuts to identify BH candidates, we test both methods recovering the known BHs in the simulated datasets $\hyp_{1-4}$. Fig.~\ref{fig:ratePurity} shows the purity of recovered BH candidates vs the fractions of PBHs in the simulated datasets. We test the purity cut method using three different priors on the individual event parameters - uniform in $\log \pi_E$ and $\log t_E$ (Table \ref{tab:priors}), a broad normal distribution for both parameters in Table 1 of \citet{Golovich:2020acu}, and uniform priors in $t_E$ and $\pi_E$.
For both methods, arbitrary thresholds need to be used to select lenses. 
In the case of the method proposed here, one must define the threshold probability, $p_{\rm threshold}(\classBH|\D, \hyp)$. 
For the linear method of \citet{Golovich:2020acu}, one must specify the target purity when calculating the line.
To assess the impact of these two parameters, we consider two choices, namely a $p_{\rm threshold}(\classBH|\D, \hyp)$ of $0.5$ and $0.9$, and a target purity of $50\%$ and $90\%$. However, defining a threshold probability, as we will see below, does not correspond to setting the final purity of the classification analysis.

For all methods, Fig.~\ref{fig:ratePurity} shows that more BHs are recovered and the purity of each sample increases as the number of PBHs in the dataset increases. This is due to PBHs increasing the abundance of BHs which populate the lower right corner of the $\pi_{E}-t_{E}$ space \citep[see Fig. 5 in][]{Pruett2022}, making it easy to separate from other lens classes. Fig.~\ref{fig:ratePurity} also shows that the probabilistic lens class method outperforms the purity based methods across all simulated datasets as measured by the final purity of the sample. The largest performance gains are for simulated datasets with low numbers of PBHs.
While not predicting quite as many correct BH candidates, the probabilistic method does lead to far fewer false positives, as shown in the upper two panels of Fig.~\ref{fig:ratePurity}.

Fig.~\ref{fig:ratePurity} also shows the sensitivity of the purity cut methods to the prior distributions used when modeling each event. All the priors appear to be uninformative in different ways, however, the log-uniform prior selects at least an order of magnitude more BH candidates and a significantly less pure sample across all simulated datasets when compared to the prior used by \citet{Golovich:2020acu} or the uniform prior. 
This is due to smaller objects, like stars, having poorly measured $\pi_{E}$. In this case, the constraint on $\pi_{E}$ for stars is driven by the prior. 
Moreover, the purity cut methods relies on the posterior mean and not the full distribution. Overall, if the $\pi_{E}$ prior mean is in the region of $\pi_{E}-t_{E}$ space which is dominated by BHs (as the log-uniform prior is), it will bias all the events to be classified as a BH when using the purity cut method. When using the uniform or normal distribution, we see a much better performance, where these two choices generally agree.

In contrast, Fig.~\ref{fig:ratePurity} illustrates the \emph{insensitivity} of the method in this paper to the arbitrary threshold probability. 
Changing the threshold from $0.9$ to $0.5$ increased the number of candidate events, but at an almost identical purity. 
The additional candidates gained by changing the threshold were equally likely to be a BH as a star.
This comes from the distribution of $p(\classBH| \D,\hyp)$ for each survey, where the integral of this distribution ultimately determines the purity, not the lower boundary.
On top of this insensitivity, we also note that this method is independent of the priors used when analyzing single events (see Eq.~\eqref{eq:finalPosteriorclassIS}), removing a possible source of systematic bias.

\begin{figure}
\includegraphics[width=\linewidth]{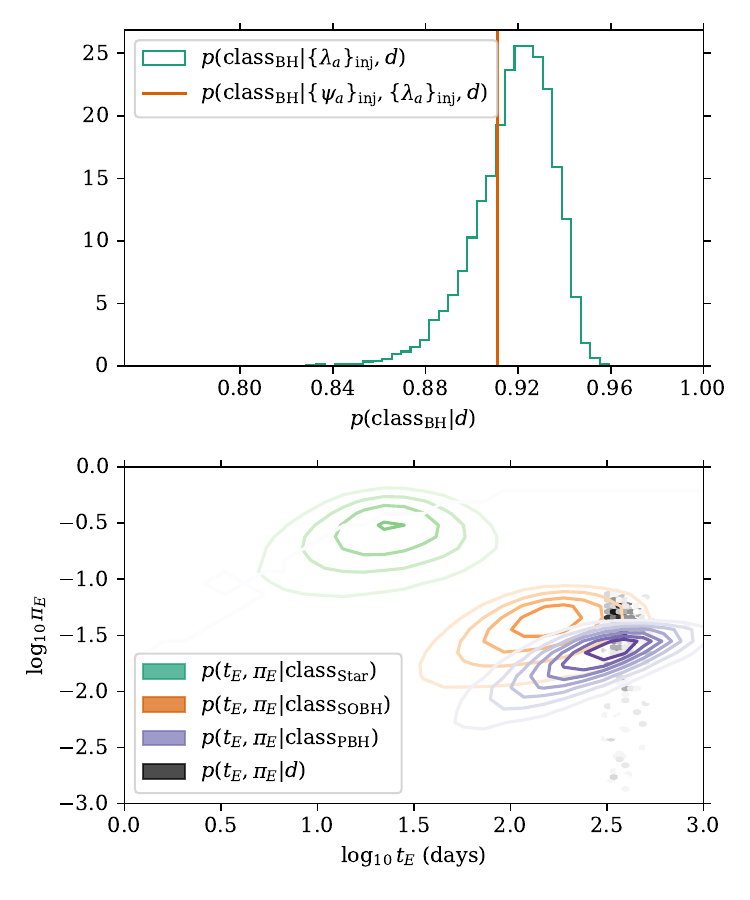}
\caption{
	\textbf{Top}: the probability distribution of a specific event to involve a BH lens (from $\hyp_2$, in this case), marginalized over the posterior distributions of mixing fractions for the underlying population model (assuming fixed shape parameters $\{\hypshapea\}_{\rm inj}$ matching those used to create the data).
	The vertical line represents the probability for the event to involve a BH lens performing the analysis with the entire population model used to create the data (assuming both shape parameters and abundances fixed to those of the model which created the data, $\{\hypshapea\}_{\rm inj}$ and $\{\popfraca\}_{\rm inj}$).
	\textbf{Bottom}: the posterior of the event in question (shown in black) as compared to the predicted distributions of each subpopulation in $t_E$-$\pi_E$ space, for comparison.
	This particular event has a high probability of being a BH lens because of its large overlap with the SOBH and PBH distributions, despite uncertainty on the expected contribution of each subpopulation to the overall lens population.
    Critically, the bottom panel of this figure only represents a graphical representation of the \emph{likelihood}.
    To determine the total probability a lens belongs to a certain class, one must also incorporate the prior probability.
    The contours are linearly spaced between 0 and 3.5.
}\label{fig:singleEventProbComp}
\end{figure}

The above tests assume that we know the underlying true lens population model, which in reality is not true. To mitigate this, our probabilistic lens classification method can marginalize over a set of possible population models. In this case, instead of a point estimate for probability of the lens classification, we have a distribution of possibilities which captures the underlying lens population uncertainty. Fig.~\ref{fig:singleEventProbComp} shows the distribution of $p(\classBH|\{\hypshapea\},\D)$ for a single event obtained when marginalizing over a set of restricted population models that only allow lens subpopulations mixing fractions to vary (see App.~\ref{app:restricted}). Fig.~\ref{fig:singleEventProbComp} shows that we were able to estimate the true $p(\classBH|\{\hypshapea\},\D)$ accurately despite not perfectly knowing the underlying relative abundances of the different subpopulations. For illustration purposes, the bottom panel shows a graphical representation of the predicted distribution in $t_E$-$\pi_E$ space for each subpopulation and the posterior of the specific event.

Two features in the results of this section provide compelling evidence for why the hierarchical analysis (results discussed in the next section) should incorporate the probabilistic nature of classification proposed in this work. 
First, our classification method never captures all the BHs in the data, as illustrated by the green band always falling below the markers representing the true number of BHs in each dataset in Fig.~\ref{fig:ratePurity}.
The standard classification schemes and the formalism of this work either miss a large fraction of BHs in the data (with a high purity) or include many BHs but accompanied with many false positives (giving a low purity).
This risks two types of bias when considering hierarchical inference: neglecting an important subset of BHs in the data or biasing results by including incorrectly classified events.
Furthermore, the majority of the results in this section were achieved by conditioning on a specific population model. 
The impact of this choice is shown in Fig.~\ref{fig:singleEventProbComp}, where the spread on classification confidence for this event can change an appreciable amount based on the uncertainty in the population model.
This illustrates the need to jointly infer the lens type along with the entire population model, simultaneously.
In the next section, we demonstrate how our methods for hierarchical inference robustly address these issues and provide unbiased results.


\section{Application to Populations}\label{sec:populationInference}

To understand the lensing population model, the results of our analysis are broken down into several parts.
We first consider the PBH subpopulation in the context of both its hyperparameter posteriors and through Bayes' factors.
We then move on to consider the stellar and SOBH lens subpopulations.

\subsection{PBH posterior information}

\begin{figure*}
\includegraphics[width=.48\linewidth]{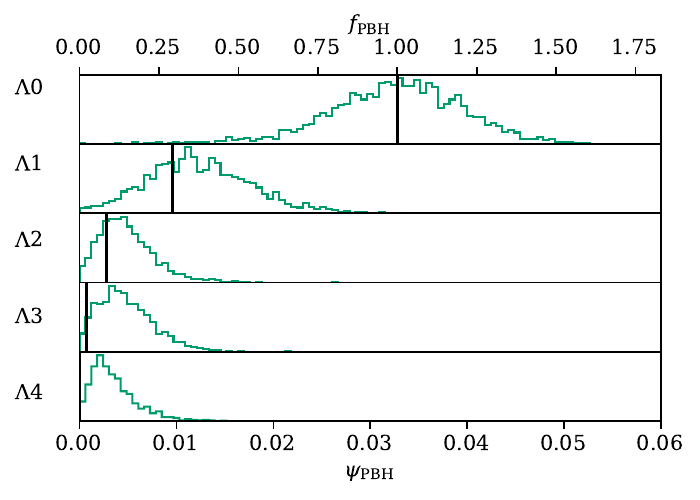}
\includegraphics[width=.48\linewidth]{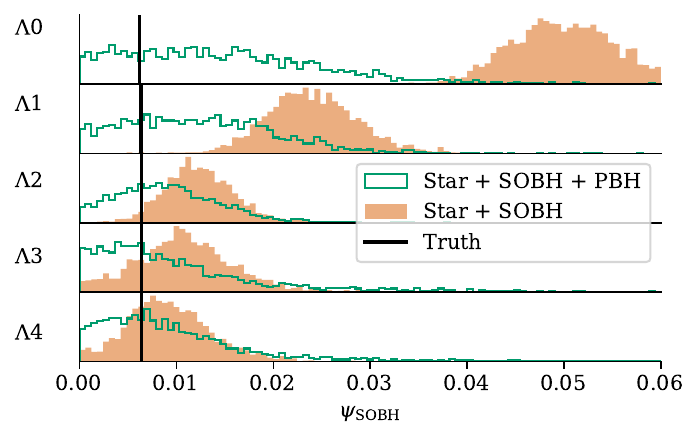}
\caption{
	Shown above are the posterior distributions obtained from out simulated datasets for the mixing fraction or relative abundance of the PBH ($\fracC$ left) and SOBH ($\fracB$ right) subpopulation.
    In the case of the PBH subpopulation, we also show its contribution to DM ($f_{\popC}$).
	Vertical lines indicate the true value used to create the data.
	The prior on the absolute abundance of each subpopulation, $N_{\classa}$, was uniform between $1$ and twice the number of lenses in the specific simulation, which translated to a prior on the relative abundance that was broad (between $0$ and $1$) but mildly peaked around $0.5$.
	For the PBH subpopulation, we see an indication that the PBH subpopulation contributes meaningfully to the explanation of the data, due to the posteriors being inconsistent with zero in the case of datasets $\hyp_0$ and $\hyp_1$ and mildly contributes in the case of $\hyp_2$.
	In the case of the other datasets ($\hyp_3$ and $\hyp_4$), we can merely make statements about the maximum contribution of PBHs to the lensing population ($\sim1\%$) and their contribution to the DM fraction ($\sim 20-25\%$).
    While datasets $\hyp_{3-4}$ peak slightly away from zero, the mean of the distribution is less than $1.5\sigma$ and is due to correlations with the SOBH abundance and the marginalization process.
	For the SOBH subpopulation, we see a separation of the posteriors when using two and three subpopulations.
	This indicates that including the PBH component leads to a better description of the data, showing preference for the more complex model.
    The (Star + SOBH) model is not flexible enough to explain the data when considering $\hyp_0$, $\hyp_1$, $\hyp_2$.
}\label{fig:fractionRidge}
\end{figure*}

We begin by evaluating our ability to measure the relative abundances of the different subpopulations, focusing first on PBHs. Fig.~\ref{fig:fractionRidge} shows that our ability to detect PBHs varies with the number of PBHs in the data. When a significant number of PBHs exist in the data ($\hyp_{0}$), we recover an accurate $\fracC$ posterior inconsistent with zero (${>}4.5 \sigma$), providing evidence for a PBH subpopulation. As the number of PBHs in the simulated data are stepped down, we find strong ($\hyp_{1}$), then mild ($\hyp_{2}$), and finally no ($\hyp_{3,4}$), constraint with the recovered $\fracC$ posterior being inconsistent with zero. As the number of PBHs in the data decreases, our ability to measure a non-zero PBH abundance diminishes. However, even in the case of $\hyp_{3,4}$, we can still place an upper bound on the PBH subpopulation, which in this case, plateaus at $\approx20$-$25\%$ of the DM fraction ($f_{\popC}$) and at $\approx1\%$ of all lenses in the population ($\fracC$).

The mixing fraction of SOBHs, $\fracB$, also encodes information about the existence of a subpopulation of PBHs. Fig.~\ref{fig:fractionRidge} shows the $\fracB$ posterior for both the two (Star + SOBHs) and three (Star + SOBHs + PBHs) population configuration. When a substantial number of PBHs exist in the data ($\hyp_0$) and only the Star + SOBHs population model is used, the SOBHs subpopulation absorbs the PBHs making $\fracB$ $\approx5$ times larger than than its true value. However, this signature of a PBH subpopulation is unlikely to be useful when applying this method to real data, due to it being completely dominated by the factor of ${\sim}100$ uncertainty on the expected number of SOBHs in the Milky Way \citep{Samland1998,1996ApJ...457..834T,1992eocm.rept...29V}.

The Star + SOBHs model cannot, however, completely absorb and explain away the PBHs in the case of a large number of PBHs actually contained in the data ($\hyp_0$ and $\hyp_1$). If the SOBH and PBH subpopulations were perfectly degenerate, the two $\fracB$ posterior distributions would be wide but overlapping, extending from ${\sim}0$ to ${\sim}0.05$ for both classes of models. This is because the sum of the PBH and SOBH subpopulations would always account for the total contribution of both subpopulations. However, because the population model favors \emph{not} having a large number of SOBH lenses but instead tends toward an SOBH subpopulation consistent with zero relative abundance when including a PBH subpopulation in the modeling, a hierarchy emerges in the explaining power of each class of population model.
In this case, using both the SOBH and PBH subpopulation to describe the entire BH subpopulation is more informative than the population model which neglects the PBH subpopulation.

\begin{figure}
\includegraphics[width=.95\linewidth]{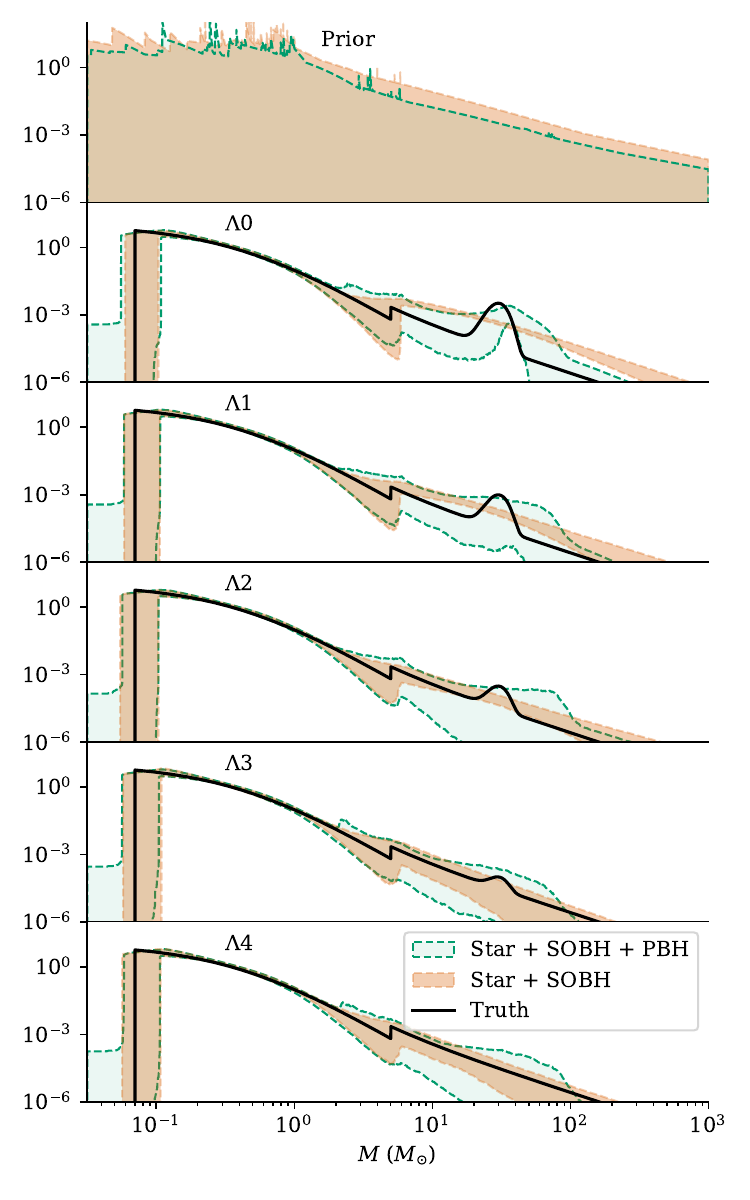}
\caption{
	Shown above are the various reconstructions of the mass spectrum of lensing objects from our ten analyses performed on five sets of data.
	The range of reconstructions allowed by the prior is shown in the top panel.
	Each subsequent panel shows the $90\%$ confidence reconstruction from the posterior probability of the inference analysis for each set of data, $\hyp_i$.
	For each dataset, we conduct two analyses: one assuming two subpopulations (hatched) and one assuming three (unhatched).
	The true distribution used to create the synthetic data is shown as a black dashed line.
	From this figure, we can see that our analysis provides an unbiased reconstruction of the underlying true distribution, accurate to within $90\%$ confidence when systematic bias is not present.
	The ability to disentangle the subpopulations is clear when a large subpopulation of PBHs is present in the data ($\hyp_0$), but these conclusions become increasingly more uncertain as the ``strength'' of the PBH subpopulation shrinks ($\hyp_1$ to $\hyp_4$).
}\label{fig:massReconRidge}
\end{figure}

Fig.~\ref{fig:massReconRidge} shows that our method can recover the lens mass spectrum across all lens subpopulations. For the simulated datasets $\hyp_{0-4}$ and for the two (Star + SOBHs) and three (Star + SOBHs + PBHs) population models, both the prior and posterior distributions on the lens mass spectrum are shown. In all cases, when the true model that generated the simulated data is used, we recover an unbiased lens mass spectrum in agreement with the true distribution to within $90\%$ credibility. The disagreement between the two classes of populations models is greatest for the datasets with the most PBH lenses. Without the flexibility of the PBH subpopulation component, the SOBH power law subpopulation shifts to compensate for the missing category of lenses. This leads to fine-tuning of the SOBH mass spectrum in the population model that neglects the PBH subpopulation because only a narrow part of the SOBH subpopulation parameter space yields reasonable agreement with the data. This effect can been seen in Fig.~\ref{fig:massReconRidge} as a narrowing of the SOBH mass spectrum between $10-10^{3}M_{\sun}$ for $\hyp_{0,1}$.

Fine tuning is an aspect of model complexity that must be considered when evaluating competing models. For detecting a subpopulation of PBHs, we have a more flexible population model (Star + SOBHs + PBH) that we have to compare against less flexible population model (Star + SOBHs) that requires fine tuning to explain the data. Overall, Fig.~\ref{fig:massReconRidge} provides a diagnostic in the model selection problem of determining the evidence for the additional subpopulation of PBH lenses, where we see a systematic inability of the simpler model to accurately recover the true distribution.

As the number of PBHs in the data decreases (from $\hyp_1$ through $\hyp_4$), our ability to disentangle the subpopulation mass spectra drops. This is shown in Fig.~\ref{fig:massReconRidge}  by the similar mass spectrum reconstructions between the two and three subpopulation models. This suggests that the extra flexibility of the higher dimension model is unwarranted or is fitting the noise. In the case of $\hyp_{1,2}$, there is still marginal evidence for the existence of the PBH subpopulation, although this is difficult to claim based purely on the mismatch between the posterior mass spectra between the two subpopulation models.  

\begin{figure}
\includegraphics[width=\linewidth]{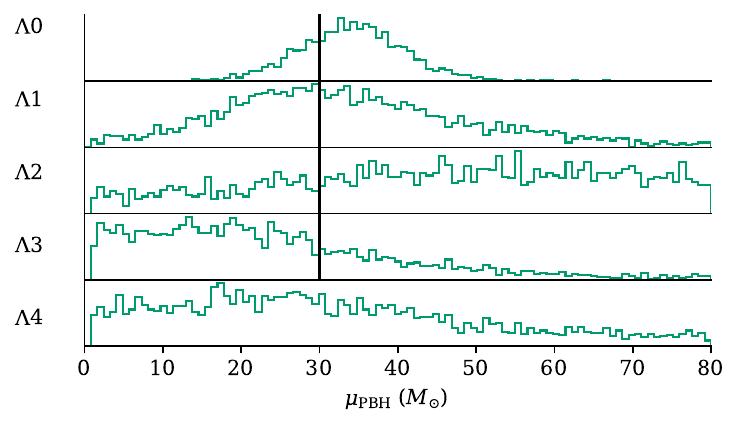}
\caption{
	Shown above are the various posterior distributions on the location parameter for the PBH Gaussian bump, $\mu_{\popC}$.
	The true values used to create the data are shown as solid vertical lines.
	For reference, the prior for this parameter was uniform between $1 M_{\odot}$ and $80 M_{\odot}$.
	For $\hyp_0$ and $\hyp_1$, the posteriors favor the true value.
	The other datasets are less informative, simply ruling out a high-mass component of the lensing population.
	The lower mass parts of the spectrum overlap with the stellar subpopulation (which dominate the catalog by a large margin), allowing the extra flexibility to be absorbed by this primary subpopulation.
}\label{fig:PBHLocationRidge}
\end{figure}

Fig.~\ref{fig:PBHLocationRidge} shows the posterior constraints on $\mu_{\popC}$ and shows that for $\hyp_{0,1}$ information can be inferred about the structure of the PBH mass spectrum. We see strong evidence for the PBH mass spectrum bump around the correct location of $30 M_{\odot}$.  The recovered PBH mass spectrum bump is always wider than true values, which indicates that our method is not sensitive to the width of the PBH mass spectrum bump compared to the location of its peak.  For $\hyp_{2-4}$, Fig.~\ref{fig:PBHLocationRidge} shows no constraint on $\mu_{\popC}$, therefore the data did not favor a high-mass PBH component in the lensing population, and only an upper bound can placed on the mass range of the PBH subpopulation. In these cases, the lower part of the PBH mass spectrum overlaps with the dominating stellar subpopulation, which can absorb the PBHs as noise.

\begin{figure}
\includegraphics[width=\linewidth]{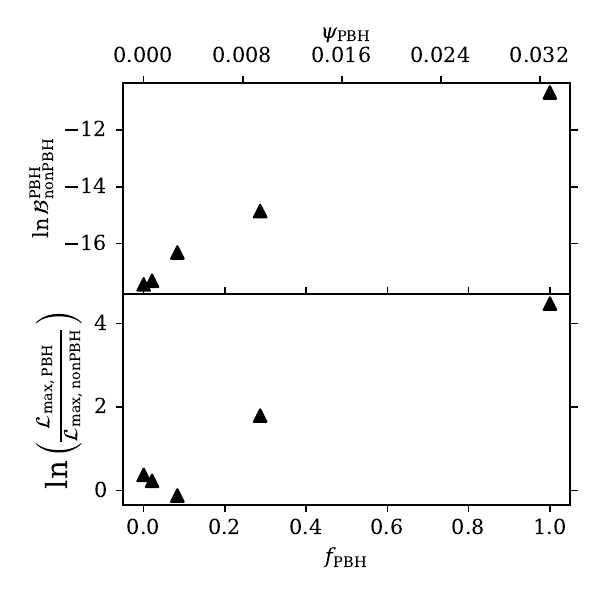}
\caption{
	\textbf{Top}: the logarithm of the Bayes' factor between the two and three subpopulation class of models for each set of data $\hyp_i$.
	\textbf{Bottom}: the logarithm of the ratio of the maximum likelihood values between the two and three subpopulation class of model for each set of data $\hyp_i$.
	While the Bayes' factor statistic does not suggest strong evidence for a PBH subpopulation in any dataset, the ability to differentiate between the two models (two or three subpopulations) clearly improves as the number of PBH lenses actually contained in the data increases (from $\hyp_4$ to $\hyp_0$).
}\label{fig:bayesFactorLikelihoodComp}
\end{figure}

\subsection{Evidence for a simulated PBH subpopulation}

In addition to examining the posteriors of the two subpopulation models, we can also compare their overall performance on the datasets directly. There are many statistics that can be used to compare competing models which all have their advantages and drawbacks. From $\chi^{2}$-based metrics \citep[e.g.,][]{2016MNRAS.458.3012W, Andrae2010}, to information criteria \citep[e.g.,][]{Kains2018} to Bayes' factors \citep[e.g,][]{Jenkins2011} and cross validation scores \citep[e.g.,][]{Welbanks2023,2023MNRAS.520..259M}, these statistics can estimate and approximate different aspects of model performance. Here we compare models using the maximum likelihood and the Bayes' factor, where the Bayes' factor is estimated as a byproduct of the parallel tempering MCMC methods described in Section \ref{sec:pop_realization}. The Bayes' factor is a widely used method for model selection and its advantages include its interpretation as the comparison of the posterior probability for each model, and that it penalizes model complexity not supported by the data. The Bayes' factor's main draw back is its sensitivity to prior distributions. Despite the Bayes's factor not being the perfect model comparison tool, we find it informative for our problem.

Fig.~\ref{fig:bayesFactorLikelihoodComp} shows that for all simulated datasets the Bayes' factor always disfavours a PBH subpopoulation. This is driven by the wide, uninformative priors used in all models (Sec.~\ref{sec:verification}), and that the two subpopulation model can partly absorb the PBH subpopulation with some fine tuning, as discussed above. While sufficient evidence cannot be found for any of the data sets in isolation with the Bayes' factor, there is a strong trend in Fig.~\ref{fig:bayesFactorLikelihoodComp} showing that as the number of PBH lenses in the data increases, our ability to distinguish between the two subpopulation model classes improves significantly. The difference in the logarithm of the maximum likelihood increases by $\approx4-5$ (equivalent to the maximum likelihood value of the three subpopulation model increasing by a factor of $\sim100$) and the logarithm of the Bayes' factor increases by $\sim 7$. While conclusive evidence for a subpopulation of PBH lenses cannot be claimed in this toy example with the Bayes' factor, the trend of its improvement from $\hyp_{0-4}$ suggests that given sufficiently informative priors it could be used to determine the presence of a PBH subpopulation. 

\begin{figure*}
\includegraphics[width=.48\linewidth]{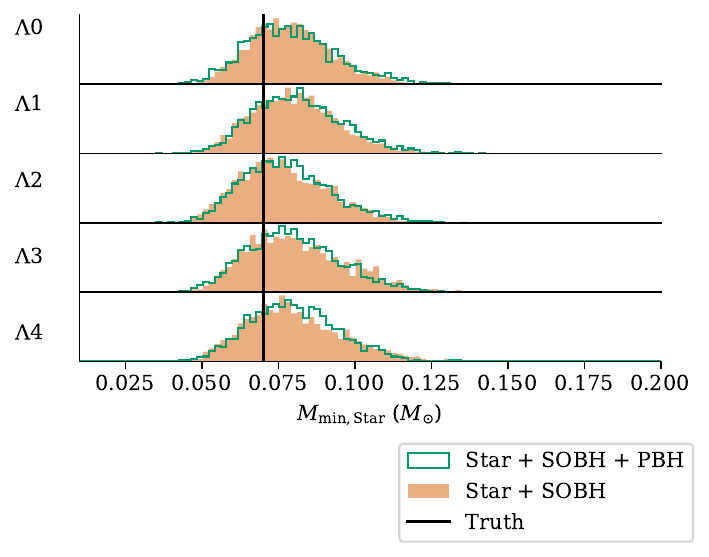}
\includegraphics[width=.48\linewidth]{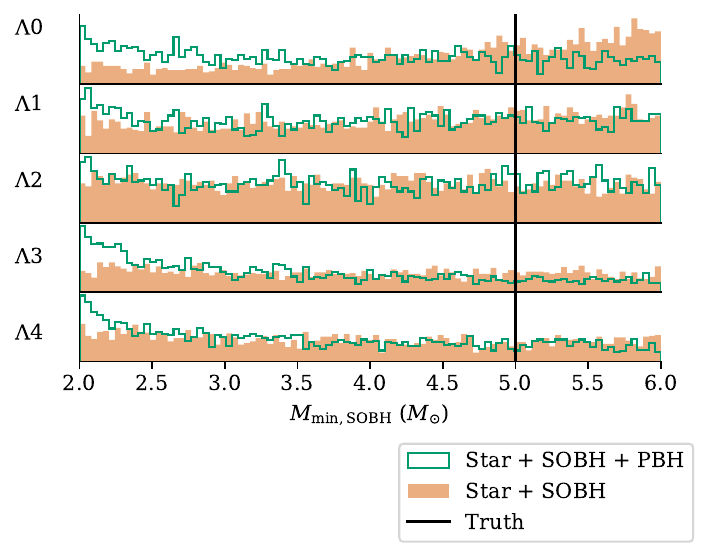}
\caption{
	Shown above are the posterior distributions for the minimum mass of the stellar (left) and SOBH (right) power law distribution ($M_{\rm min, \popA}$ and $M_{\rm min, \popB}$).
	The vertical lines indicate the true value used to create each dataset, while the hatched distributions refer to inference performed assuming two subpopulations and unhatched distributions were inferred assuming three subpopulations.
	The prior for $M_{\rm min, \popA}$ is a uniform distribution between $0.01 M_{\odot}$ and $1 M_{\odot}$.
	The prior for $M_{\rm min, \popB}$ is a uniform distribution between $2 M_{\odot}$ and $6 M_{\odot}$.
	We see that we can robustly measure the minimum mass of the stellar power law distribution, regardless of the data set or population model employed.
	However, we \emph{cannot} infer the minimum mass of the SOBH power law distribution, regardless of the data set or population model employed.
	The data sets and type of data (photometry only) do not contain enough information to place robust bounds on the minimum mass of the SOBH subpopulation that would be useful in measuring things like the NS-BH mass gap~\citep{2011ApJ...741..103F,2010ApJ...725.1918O,2001ApJ...554..548F}. 		
}\label{fig:MinMassRidge}
\end{figure*}

\subsection{Stars and SOBHs}

Beyond PBHs, we can asses the ability of the model to infer features of the SOBH and stellar mass spectra. Fig.~\ref{fig:MinMassRidge} shows the posterior constraints on $M_{\rm min, \popA}$ and $M_{\rm min, \popB}$. We find that for all population models and simulated datasets we are able to obtains a tight constraint on $M_{\rm min, \popA}$ of roughly $0.08 M_{\odot}\pm 0.02 M_{\odot}$. The large number of stellar lenses in the data suggests there is enough information to make robust claims about the minimum stellar mass. Conversely, we find $M_{\rm min, \popB}$ is never constrained due to its posterior distribution always approximately recovering the prior distribution. This suggests we are not able to probe the existence and properties of the mass gap between NSs and BHs~\citep{2011ApJ...741..103F,2010ApJ...725.1918O,2001ApJ...554..548F}. While this conclusion should be revisited for future surveys with tighter measurements and larger catalog sizes ~\citep[e.g., Roman Space Telescope;][]{2015arXiv150303757S} or when other data is taken into account (such as astrometry), our initial analysis using photometric only, OGLE-type data does not give confidence that this will be a measurable feature.

\begin{figure*}
\includegraphics[width=.48\linewidth]{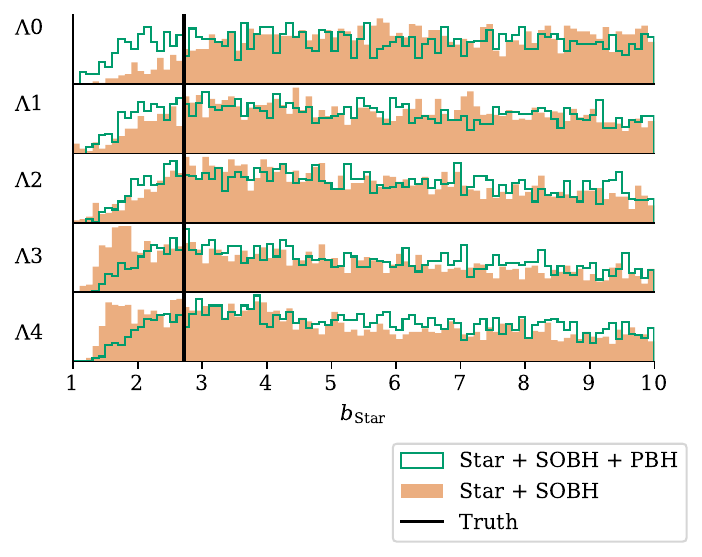}
\includegraphics[width=.48\linewidth]{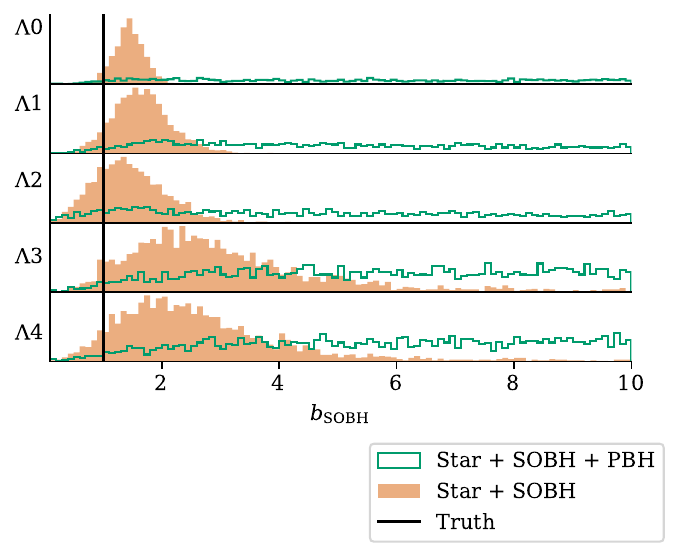}
\caption{
	Posterior distribution for the shape parameter of the stellar (left) and SOBH (right) power law distribution ($b_{\popA}$ and $b_{\popB}$).
	The vertical lines indicate the true value used to create the data.
	The two types of analyses (assuming two or three subpopulations) are shown as a hatched and non-hatched distribution, respectively.
	The prior for these parameters is a uniform distribution between $0.1$ and $10$.
	For the stellar subpopulation, the posterior is approximately the prior distribution because of correlations with the location parameter, $\mu_{\popA}$.
	This correlation is shown in Fig.~\ref{fig:bmuStarCov}.
	Considering the SOBH subpopulation, the posteriors are uninformative in the case of assuming three subpopulations.
	However, when only considering two subpopulations (thereby focusing on astrophysics as opposed to exotic physics), the posteriors contain significantly more information than the priors.
    The slight bias in the posteriors are connected to the same correlation shown in Fig.~\ref{fig:bmuStarCov}, but for the SOBH subpopulation model, and the process of marginalization.
}\label{fig:PowerRidge}
\end{figure*}

\begin{figure}
\includegraphics[width=\linewidth]{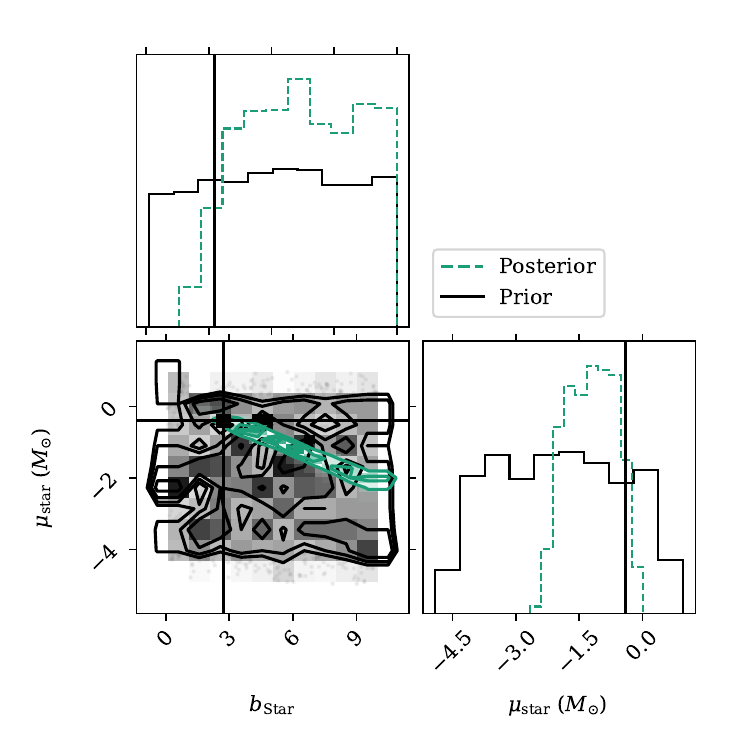}
\caption{
	The two dimensional joint posterior on $b_{\popA}$ and $\loc_{\popA}$, as inferred using a two subpopulation model and the $\hyp_0$ data set.
	The green (dotted) contours and histograms refer to the posterior, while the black (solid) contours and histogram refer to the prior distribution for these parameters.
	The solid black lines represent the true values of the parameters used to generate the data.
	The strong correlation causes the one dimensional, marginalized posteriors on $b_{\popA}$ and $\mu_{\popA}$ to be very broad, despite there being plenty of information about certain linear combinations of these parameters in the data.
}\label{fig:bmuStarCov}
\end{figure}

Fig.~\ref{fig:PowerRidge} shows the posterior constraints on the stellar and SOBH subpopulations shape parameters $b_{\popA}$ and $b_{\popB}$, respectively. The posteriors on $b_{\popA}$ generally match those of the prior. This initially seems surprising given the tight reconstruction of the stellar mass spectrum in Fig.~\ref{fig:massReconRidge}. However, the broad posteriors for $b_{\popA}$ are caused by the correlation with the location parameter, $\loc_{\popA}$, which is a symptom of the flexible Pareto type-II distribution model used. $\loc_{\popA}$ smooths the low mass component of the distribution, resulting in a high degeneracy with the shape parameter $b_{\popA}$. Fig.~\ref{fig:bmuStarCov} shows this effect - a certain linear combination of $b_{\popA}$ and $b_{\popA}$ is constrained tightly, while the orthogonal combination is almost totally unconstrained. 

Finally, Fig.~\ref{fig:PowerRidge} shows the posteriors for the shape parameter for the SOBH subpopulation, $b_{\popB}$. When the Star + SOBH + PBH population model is fit, the $b_{\popB}$ posteriors are largely uninformative. However, when fitting the Star + SOBH population model, informative posterior constraints on $b_{\popB}$ are possible. For the simulated data sets with a high number of PBHs (e.g., $\hyp_{0}$) we see a systematic error in the $b_{\popB}$ posterior caused by the model-simulation miss match. However, in the case of no systematics between the model and the simulated dataset ($\hyp_4$), we see a reasonably tight constraint on the SOBH shape parameter.


\section{Conclusions and Future Work}\label{sec:conclusions}

In this work, we proposed and validated a methodology to conduct hierarchical inference simultaneously with the lens classification for individual events.
The benefits of this framework over existing methods include properly accounting for Poisson statistics, measurement uncertainty, and selection bias while not assuming definite criteria for lens classification. 
This framework allows marginalization over population uncertainty, which is key to reliable inference given the current uncertainty about underlying lens population characteristics.

On a single event level, our method outperformed current purity-cut strategies used to classify lenses in search of BH candidates. We were able to recover purer samples of BH events while also quantifying population model uncertainty. Our probabilistic lens classification scheme also revealed and quantified intrinsic degenerate structure in the $t_{E}-\pi_{E}$-space. Further investigation of this structure could yield insights into reliably identifying BH microlensing events in real-time photometrically and efficiently allocating astrometric followup resources \citep[e.g.,][]{Lu2016,Sahu2022,2022ApJ...933L..23L}. Although photometric microlensing parameters are difficult to constrain early in an event's evolution \citep{Albrow2004}, our classification method could be used to classify an event at or after its photometric peak, where $t_{E}$ and $\pi_{E}$ are better constrained, to decide whether to use astrometric resources to measure the second astrometric peak. \citep[e.g.,][]{Dominik2000}. Further work on how well microlensing parameters can be constrained from a partial lightcurve would also benefit classification efforts \citep[e.g.,][]{Dominik2009}. Real-time identification of black hole microlensing events is likely to become more important in the era of Rubin LSST survey planing \citep[e.g.,][]{Street2023} and the integration of automated identification and followup planning methods into Target and Observation Management systems \citep[e.g.,][]{Street2018, Coulter2023}

our classification method could be used to classify an event at or after its photometric peak, where $t_{E}$ and $\pi_{E}$ are better constrained, to decide whether to use astrometric resources to measure the second astrometric peak. Further work on how well the microlensing parameters can be constrained from a partial lightcurve would be fruitful

We find that our full hierarchical model leads to inference on the lens population mass spectrum and abundances that is accurate and effective while appropriately handling population uncertainty. In the context of characterizing a PBH subpopulation of lenses, our method produces posterior constraints on the PBH abundance inconsistent with zero at ${>}4.5\sigma$ when considering $f_{\rm PBH}=1$. For the more realistic case of $f_{\rm PBH}\lesssim 0.25$ our ability to identify the subpopulation begins to deteriorate, and we are only capable of placing upper limits on PBH subpopulation. Moreover, a PBH subpopulation signature for any $f_{\rm PBH}$ will likely be derived through a joint analysis of maximum likelihood measurements, Bayes' factors, and hyperparameter posteriors.

The results here can be compared to the constraints from past collaborations \citep[e.g.,][]{Macho:2000nvd,EROS-2:2006ryy,2009MNRAS.397.1228W,2010MNRAS.407..189W,2011MNRAS.413..493W,2011MNRAS.416.2949W,2022A&A...664A.106B} which claim constraints on the DM fraction of MACHOs in the range of $\approx2-20\%$ for different mass ranges between $10^{-7}{-}10^3$ by studying events towards the Magellanic Clouds. However, direct comparison to the original microlensing MACHO constraints will have to be performed carefully due to two main reasons. Firstly, the effects of systematic noise arising for events detected towards Bulge (PBH confusion with SOBH and Stellar lenses) vs the Magellanic clouds \citep[Galactic disk and self-lensing;][]{2011MNRAS.416.2949W} are different. Secondly, the method presented in this work jointly infers a PBH mass spectrum and abundance which will have to be reconciled with the classic restrictive model of a delta mass function - a distinction in modeling assumptions shown to be important \citep{Green2016}. Although modeling the PBH mass spectrum complicates comparison with the original Magellanic Cloud MACHO constraints, this key improvement is more general and will allow microlensing to contribute to other ongoing studies of the dark mass spectrum \citep[e.g.,][]{LIGOScientific:2020kqk,LIGOScientific:2021psn,Zevin:2020gbd,Franciolini:2021tla}.

For stellar lenses, the population constraints derived were largely independent of the number of BHs in the data, due to stellar lenses vastly outnumbering BH lenses. For SOBH lenses, the slope of the power law mass spectrum and their relative abundance can be accurately constrained if a PBH subpopulation is not present in the data. When PBHs were injected into the data, degeneracies between SOBH and PBH subpopulation models made it difficult to disentangle the characteristics of the two subpopulations. In all cases, we found the the minimum SOBH mass difficult to constrain. This suggests that extracting information about a possible SOBH mass gap via microlensing \citep[e.g.,][]{2020A&A...636A..20W} is likely difficult with current photometric surveys and will require more data (through larger and/or longer surveys), increased measurement precision or the incorporation of further information such as astrometric microlensing information. 

There are multiple avenues of future research to be taken.
The most immediate would be applying these methods to OGLE-IV data, beginning with the context of better understanding the BH population.
To accomplish that, the simple population model presented here will need to be expanded to include more realistic information, such as the distributions of the flux blending fraction, the velocity distributions, the spatial distributions, other subpopulations like neutron stars and white dwarfs, etc. 
Implementing these extensions is purely a practical concern, as they can be formally integrated into the analysis through the framework presented here.
These methods can also be extended to better understand other subpopulations of lenses.
Variations have already been applied in the context of free floating planets~\citep{2023arXiv230308280S}, but the framework outlined here can help to improve those constraints by marginalizing over other population uncertainties, accounting for Poisson statistics and more carefully accounting for selection bias.

Finally, we also note that this methodology can easily be extended to heterogeneous data, i.e., incorporating simultaneous astrometric observations \citep[e.g., for the Roman Space Telescope;][]{Sajadian2023, Lam2023} or follow-up astrometric measurements from current space telescopes \citep[e.g.,][]{Sahu2017,Kains2017,Zurlo2018,2022ApJ...933L..23L,Sahu2022,2023MNRAS.520..259M}.
The integration of both types of data will prove to be indispensable as they probe different event parameters and different distributions of events in the galaxy and can break photometric microlensing degeneracies.
In the case of current astrometric measurements, the low number of events being followed up suggests a small impact to hierarchical inference.
However, when considering future surveys like the Roman Space Telescope, the impact of this joint analysis remains an open question and should be investigated thoroughly.

\begin{acknowledgments}
This work was performed under the auspices of the U.S. Department of Energy by Lawrence Livermore National Laboratory under Contract DE-AC52-07NA27344. The document number is \IMRELEASENO{}. This work was supported by the LLNL-LDRD Program under Project 22-ERD-037. 
M.F.H. acknowledges the support of a National Aeronautics and Space Administration FINESST grant under No. ASTRO20-0022.
C.Y.L. and J.R.L. acknowledge support from the National Science Foundation under grant No. 1909641 and the Heising-Simons Foundation under grant No. 2022- 3542. 
C.Y.L. also acknowledges support from NASA FINESST grant No. 80NSSC21K2043, a research grant from the H2H8 Foundation, and a Carnegie Fellowship.
The authors would like to thank Alex Geringer-Sameth, Michael Schneider, James Barbieri and James Buchanan for helpful discussions. 
\end{acknowledgments}

\software{This research has made use of NASA's Astrophysics Data System Bibliographic Services. NumPy \citep{Harris2020}, SciPy \citep{2020SciPy-NMeth}, Matplotlib \citep{Hunter2007}, corner \citep{corner}, emcee \citep{Foreman-Mackey2013}, scikit-learn \citep{sklearn_api}, Singularity \citep{Kurtzer2017, kurtzer2021}, Docker \citep{merkel2014docker}.}

\appendix

\section{Hierarchical Derivation}\label{app:hierarchicalDeriv}

In this appendix, we summarize some past work on hierarchical inference with astronomy data while incorporating measurement uncertainty and selection bias~\citep[e.g.,][]{Loredo:2004,Vitale:2020aaz,Mandel:2018mve,Taylor:2018iat}.
We outline the calculation conditioning on detection throughout, as the authors found this helped dispel misconceptions about hierarchical studies with observation bias.
The key idea behind observation bias is that it enters via the selection process. 
This means that events in the universe that were not detected must be explicitly accounted for through marginalization.
Without this detail, selection bias cannot be accurately account for.

Generally, we would like to obtain the posterior distribution on $\hyp$ given we have $\NOBS$ data segments $\{\DDi\}$ (lightcurves in the case of microlensing events) all containing a trigger (for $i\in[0,\NOBS)$) and set of triggers for all segments of data $\{\trig\}$,
\begin{equation}
p(\hyp | \{\DDi\}, \{\trig\},\NOBS)= \frac{p(\hyp) p(\{\DDi\}, \{\trig\}, \NOBS| \hyp) }{p(\{\DDi\}, \{\trig\}, \NOBS)}\,,
\end{equation}
where we have the population prior, $p(\hyp)$, the population likelihood, $p(\{\DDi\}, \{\trig\}, \NOBS | \hyp)$, and the population evidence, $p(\{\DDi\}, \{\trig\}, \NOBS )$.

However, this only accounts for the data we have identified as containing a trigger. 
In order to account for detected and non-detected signals, we will include all the data into the likelihood term and marginalize over all the parameters associated with those non-trigger segments of data.
Furthermore, we will assume all events with a trigger are truly microlensing events, meaning no background events from noise or other artifacts induce a trigger.
Explicitly, the data can be partitioned into three separate categories (given our assumptions that there's no background events): (1) segments that do contain a trigger $\{\DDi\}$ with associated triggers $\{\trigdi\}$ for $i \in [0,\NOBS)$ for $\NOBS$ segments, (2) segments that do \emph{not} contain a trigger but \emph{do} contain an event, $\{\DNDj\}$ with associated \emph{non}-triggers $\{\trigndj\}$ for $j \in [0,\NNOBS)$ for $\NNOBS$ segments and (3) segments that do \emph{not} contain a trigger and do \emph{not} contain an event, $\{\DNEk\}$ with associated \emph{non}-triggers $\{\trignek\}$ for $k \in [0,\NNOEVENT)$ for $\NNOEVENT$ segments.
The total number of signals in the data (detected or not) is $\Nt=\NOBS + \NNOBS$.
The triggers are related by $\{\trig\} = \{\trigdi\}\cup \{\trigndj\}\cup \{\trignek\}$.

We note that we will \emph{not} marginalize over the \emph{non}-triggers themselves, as these hypotheses are in the posterior through $\{\trig\}$. 
The existence of data without triggers is known to us as observers, and we can account for that.
However, the data for those non-triggers will \emph{not} be analyzed in reality, so we will need to marginalize over those segments $\{\DNDj\}$ and $\{\DNEk\}$.
Furthermore, we do not know how many true signals were not detected, $\NNOBS$, so we will need to marginalize over that parameter as well.
Finally, we do not need to marginalize over $\NNOEVENT$, as this is not a free parameter because it is specified by the combination of $\NOBS$ and $\NNOBS$.

This gives
\begin{equation}
p(\{\DDi\}, \{\trig\} , \NOBS| \hyp) =  \sum_{\NNOBS = 0}^\infty \int d\{\DNEk\} \int d \{\DNDj \} p(\{\DDi\}, \{\trigdi\},\{\DNDj\}, \{\trigndj\} ,\{\DNEk\}, \{\trignek\}, \NOBS, \NNOBS, \NNOEVENT| \hyp)\,.
\end{equation}

We note that summing over $\NNOBS$ to $\infty$ is an approximation. Technically, because of our assumption that signals do not overlap, this is over all possible chunks of data without a trigger $\Nt - \NOBS$. Including the appropriate limits gives an extra factor of gives an extra factor of $\Gamma(\Nt-\NOBS+1, (1-\alpha)N(\hyp))/\Gamma(\Nt-\NOBS+1)\approx 1$ when considering $\Nt-\NOBS \gg (1-\alpha)N(\hyp)$. In words, this assumption amounts to assuming the predicted number of missed events in the data is much less than the amount of data not being analyzed. 

The population model, however, does not provide information about the data itself. 
We need to incorporate event model parameters (specific to each event) in order to connect the probability of seeing each set of data given a signal model to the population at large.
We will therefore introduce event parameters $\{\eparamdi\}$ and $\{\eparamndj\}$ for the observed and unobserved signals in the data, respectively. 
The data segments without a signal are modeled by our noise model alone, which we are here assuming to be fixed (with no free parameters).
We must then marginalize over these extra nuisance parameters.
Simultaneously, we will expand these probabilities out to emphasize the hierarchical structure.
This leaves the following distribution for the likelihood
\begin{align}\label{eq:populationLikelihood}\nonumber
p(\{\DDi\}, \{\trig\} , \NOBS| \hyp) &=  \sum_{\NNOBS = 0}^\infty \int d\{\DNEk\} \int d \{\DNDj \} \int d\{\eparamdi\} \int d\{\eparamndj\}p(\{\eparamdi\}, \{\eparamndj\}, \NOBS, \NNOBS, \NNOEVENT| \hyp) \\ 
&\times  p(\{\DDi\}, \{\trigdi\},\{\DNDj\}, \{\trigndj\} ,\{\DNEk\}, \{\trignek\}|\{\eparamdi\}, \{\eparamndj\})  \,,
\end{align}
where we have also used the fact that once a set of model parameters has been defined, the data no longer directly depends on the population model.

Until now, we have remained largely agnostic about the details of the data or statistical model (with the caveat of the assumptions outlined throughout, such as stationary noise, no background events, etc.), but now we assign an actual joint likelihood function to the event parameters and detection number distribution, $p(\{\eparamdi\}, \{\eparamndj\}, \NOBS, \NNOBS, \NNOEVENT| \hyp)$.
Considering these details (statistical independence of counts with a nonuniform rate), we will model the event rates as an inhomogeneous Poisson process~\citep[e.g.,][]{gregory_2005,2011ApJ...742...38Y,Loredo:2004,2011ApJ...742...38Y}.
First, we neglect the information that some sources and others are not, and simply quote the probability of seeing $\Nt$ sources with parameters $\{\eparamdi\}$ given a population model parametrized by $\hyp$,
\begin{equation}
p(\{\eparamdi\}, \Nt| \hyp) = \frac{e^{-N(\hyp)}}{\Nt!}N(\hyp)^{\Nt} \prod_i^{\Nt} p(\eparamdi|\hyp)\,,
\end{equation}
where $N$ is the prediction from the population model for the total number of sources.
The only modification we make to the results of past literature is to add back in the normalization $\Nt!$.
Normally unimportant, this factor will be relevant for marginalization, below.

Now, we can make the identification that certain events will be classified as triggering events, while others are not. 
Given this, we can partition the data into those that contain a trigger and those that don't by separating out the products and including a combinatorics term ${\Nt \choose \NOBS}$ to account for the different combinations of finding events to be detectable or not.
This yields 
\begin{equation}
p(\{\eparamdi\},\{\eparamndj\}, \NOBS,\NNOBS| \hyp)  \propto \frac{e^{-N(\hyp)}}{\NOBS! \NNOBS!}N(\hyp)^{\NOBS} N(\hyp)^{\NNOBS} \prod_i^{\NOBS} p(\eparamdi|\hyp) \prod_j^{\NNOBS} p(\eparamndj|\hyp)\,,
\end{equation}

With the joint probability of the source parameters and detection rates worked out, we can reinsert this quantity into the original likelihood, Eq.~\eqref{eq:populationLikelihood}.
This gives the new form of the likelihood 
\begin{equation}
p(\{\DDi\}, \{\trig\} , \NOBS| \hyp)  \propto \sum_{\NNOBS = 0}^{\infty}\frac{e^{-N(\hyp)}}{\NOBS! \NNOBS!}N(\hyp)^{\NOBS} N(\hyp)^{\NNOBS} \prod_{k=0}^{\NNOEVENT}\mathcal{L}_k^{\emptyset}\prod_{j=0}^{\NNOBS}\mathcal{L}_j^{\neg {\rm obs}}\prod_{i=0}^{\NOBS}\mathcal{L}_i^{{\rm obs}}\,,
\end{equation}
where we point out that the pairs of data segments and triggers are assumed to be independent of one another, and use the following definitions
\begin{subequations}
\begin{equation}
\mathcal{L}_k^{\emptyset}  = \int d\DNEk p(\DNEk, \trignek) \,,
\end{equation}
\begin{equation}
\mathcal{L}_j^{\neg{\rm obs}}  =\int d \DNDj\int d\eparamndj p(\DNDj, \trigndj| \eparamndj)p(\eparamndj|\hyp) \,,
\end{equation}
\begin{equation}
\mathcal{L}_i^{{\rm obs}}  = \int d\eparamdi p(\DDi, \trigdi|\eparamdi)p(\eparamdi|\hyp)\,,
\end{equation}
\end{subequations}

Taking these one at a time, we note that $\mathcal{L}_k^{\emptyset}$ is simply the likelihood of no signal in the data, and evaluates to a number, independent of any astrophysical models (event or population).
In reality, this depends on the noise model and is implied by the form of the likelihood function.
Therefore, we can neglect these terms, as they can be accounted for in the overall evidence as a normalization factor.
Looking at $\mathcal{L}_j^{\neg {\rm obs}}$, we get the following 
\begin{align}\nonumber
\mathcal{L}_j^{\neg {\rm obs}} & =\int d \DNDj\int d\eparamndj p(\DNDj|  \eparamndj)p( \trigndj|\DNDj, \eparamndj)p(\eparamndj|\hyp) \,,\\
 & = ( 1- \alpha) \,,
\end{align}
where we have used the product rule to separate out the joint likelihood in the first line.
The parameter $\alpha$ describes the integrated effect of selection bias, and is defined above in Eq.~\eqref{eq:alpha}.
When conditioning on \emph{non}-triggers, this gives the factor of $(1-\alpha)$, which is the complementary set.

Finally, we can simplify the final likelihood expression, giving 
\begin{align}\nonumber
\mathcal{L}_i^{{\rm obs}} & = \int d\eparamdi p(\DDi, \trigdi|\eparamdi)p(\eparamdi|\hyp)\,,\\\nonumber
&= \int d\eparamdi p( \trigdi|\DDi,\eparamdi)p(\DDi | \eparamdi) p(\eparamdi|\hyp)\,,\\
&= \int d\eparamdi p(\DDi | \eparamdi) p(\eparamdi|\hyp)\,,
\end{align}
where we have noted that we are \emph{not} marginalizing over $\DDi$, and we have actual data to look at (which has been unambiguously defined to have a trigger).
This means the probability of getting a trigger, given we have data $\DDi$ is one, giving the third line.

Incorporating these results, we obtain the new form of the likelihood
\begin{equation}
p(\{\DDi\}, \{\trig\} ,  \NOBS| \hyp)  \propto \sum_{\NNOBS = 0}^{\infty}\frac{e^{-N(\hyp)}}{\NOBS! \NNOBS!}   \left(N(\hyp)(1-\alpha)\right)^{\NNOBS} N(\hyp)^{\NOBS} \prod_{i=0}^{\NOBS}\mathcal{L}_i^{{\rm obs}}\,.
\end{equation}
Now, we can marginalize over the number of non-detected sources, yielding 
\begin{equation}
p(\{\DDi\}, \{\trig\} , \NOBS| \hyp)  \propto \frac{e^{-\alpha N(\hyp)}}{\NOBS! }N(\hyp)^{\NOBS} \prod_{i=0}^{\NOBS}\mathcal{L}_i^{{\rm obs}}\,.
\end{equation}
If we now reincorporate the prior and population evidence, we get our final answer for the posterior probability for a population model
\begin{subequations}\label{eq:popPostAPP}
\begin{equation}
p(\hyp | \{\DDi\}, \{\trig\},\NOBS) \propto \frac{p(\hyp)e^{-\alpha N(\hyp)} }{p(\{\DDi\}, \{\trig\}, \NOBS)}N(\hyp)^{\NOBS} \prod_{i=0}^{\NOBS}\mathcal{L}_i^{{\rm obs}}\,,
\end{equation}
\begin{equation}
\mathcal{L}_i^{{\rm obs}}  = \int d\eparamdi p(\DDi|\eparamdi)p(\eparamdi|\hyp)\,.
\end{equation}
\end{subequations}

If we are not interested in the overall rate (because of complications with detection efficiency, etc), but only care about the ``shape parameters'' $\{\hypshapea\}$ and relative abundances $\{\popfraca\}$, we can marginalize over the total number of sources analytically. If we use the prior 
\begin{equation}
p(N) \propto \frac{1}{N}\,,
\end{equation}
we get the following form for the marginal posterior distribution~\citep{Fishbach:2018edt}
\begin{subequations}\label{eq:marginalizedPopPost}
\begin{equation}
p(\{\popfraca\},\{\hypshapea\} | \{\DDi\}, \{\trig\},\NOBS) \propto \frac{p(\{\popfraca\},\{\hypshapea\})\alpha^{-\NOBS} }{p(\{\DDi\}, \{\trig\}, \NOBS)} \prod_{i=0}^{\NOBS}\mathcal{L}_i^{{\rm obs}}\,,
\end{equation}
\begin{equation}
\mathcal{L}_i^{{\rm obs}}  = \int d\eparamdi p(\DDi|\eparamdi)p(\eparamdi|\{\popfraca\},\{\hypshapea\})\,,
\end{equation}
\end{subequations}
Eq.~\eqref{eq:popPostAPP} and Eq.~\eqref{eq:marginalizedPopPost} are the final forms of the posterior used when inferring all the hyperparameters of the population model, the latter being useful when marginalizing over the total rate.


\section{Restricted Analysis}\label{app:restricted}

In addition to the full model in Sec. \ref{sec:stats}, we also find it informative to study a restricted version. In this restricted model, we exploit the fact that we are using a mixture model for lens classes and fix the parameters of the lens subpopulations $\{\hypshapea\}$, only allowing the population mixing fractions $\{\popfraca\}$ to vary.  In this case, 

\begin{equation}\label{eq:mixmodel}
p(\eparam | \{\popfraca \}, \{\hypshapea\})  = \sum_a^{\nclass} \popfraca p(\eparam | \hypshapea)\,,
\end{equation}

We will simply point out the simplifications that can be made if using the population model in Eq.~\eqref{eq:mixmodel}.
Each event likelihood can be rewritten as 
\begin{align}\nonumber
\mathcal{L}_i^{\rm obs} &= \sum \popfraca \int d\eparamdi p(\DDi|\eparamdi) p(\eparamdi | \hypshapea) \,,\\
 &= \sum \popfraca \mathcal{L}_{i,a}^{\rm obs} \,,
\end{align}
where we note that the integral in $\mathcal{L}_{i,a}^{\rm obs}$ can be pre-computed once for each event and each subpopulation, if the shape parameters $\hypshapea$ are fixed.

Similarly, the same argument can be used for $\alpha$
\begin{align}\nonumber
\alpha &= \sum \popfraca \int d\D \int d \eparam  p(\trig | \D)  p(\D | \eparam) p(\eparam |\hypshapea) \,\\
&= \sum \popfraca \alpha_a \,.
\end{align}
If $\alpha_a$ can be pre-computed for each subpopulation, these two simplifications lead to Eq.~\eqref{eq:mixingFractionPost}, allowing for drastic computational speedup at the cost of lost flexibility.

We also analytically marginalize the overall rate, $N$, by assigning the prior $p(N) \propto 1/N$ \citep[][Appendix \ref{app:hierarchicalDeriv}]{Fishbach:2018edt}. This marginalization allows us to bypass modeling $N$, which is difficult due to observing effects like weather and observing schedules. This marginalizing process is possible and frequently useful when inferring the entire population model as well, independent of the mixture model we use here. However, we would like to compare our restricted model to the most flexible version of Eq.~\eqref{eq:popPost} for our initial validation. This leads us to the posterior of the restricted model,

\begin{subequations}\label{eq:mixingFractionPost}
\begin{equation}
p(\{\popfraca\} | \{\DDi\}, \{\trig\},  \NOBS,\{\hypshapea\}) \propto \frac{ p(\{\popfraca\}|\{\hypshapea\}) \alpha^{-\NOBS}}{p(\{\DDi\}, \{\trig\}, \NOBS) } \prod_{i=0}^{\NOBS}\mathcal{L}_i^{{\rm obs}}\,,
\end{equation}
\begin{equation}
\mathcal{L}_i^{{\rm obs}}  = \sum_a^{\nclass}\popfraca \frac{p( \classa | \DDi , \hyp )}{\pi(\classa | \hyp)}\,,
\end{equation}
\begin{equation}
\alpha = \sum_a^{\nclass}\popfraca\alpha_a\,,
\end{equation}
\begin{equation}
\alpha_a  = \int d\D \int d\eparam p(\trig | \D) p(\D | \eparam)p(\eparam|\hypshapea)\,.
\end{equation}
\end{subequations}

Here, the right hand side is now independent of $N$, and we have a likelihood that enables quick computation. For $\mathcal{L}_i^{{\rm obs}}$, we have leveraged Eq.~\eqref{eq:finalPosteriorclass} and can run our classification inference on all the events once, and then reuse those probabilities by combining them as weighted sums. Similarly, because we have fixed $\{\hypshapea\}$, we can pre-compute $\alpha_a$ values which can be added in weighted sums to calculate $\alpha$ quickly. 

\begin{figure*}
\centering
\includegraphics[width=.48\linewidth]{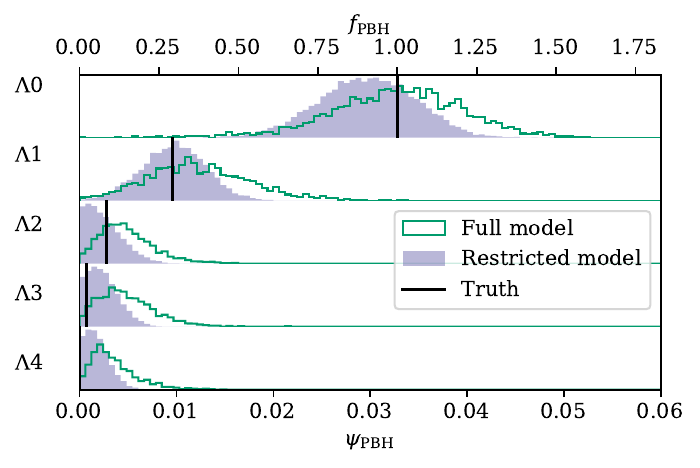}
\includegraphics[width=.48\linewidth]{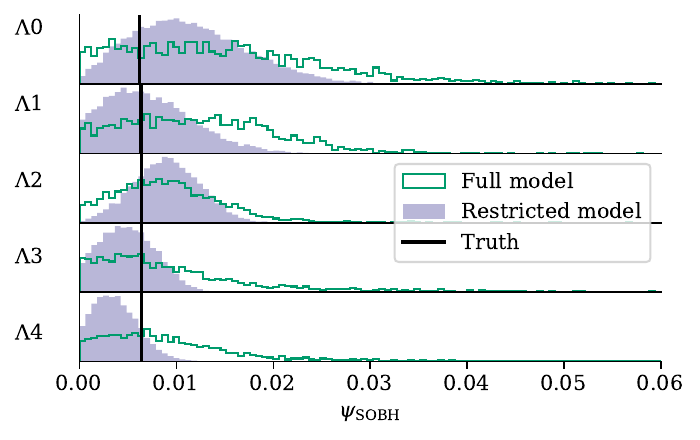}
\includegraphics[width=.48\linewidth]{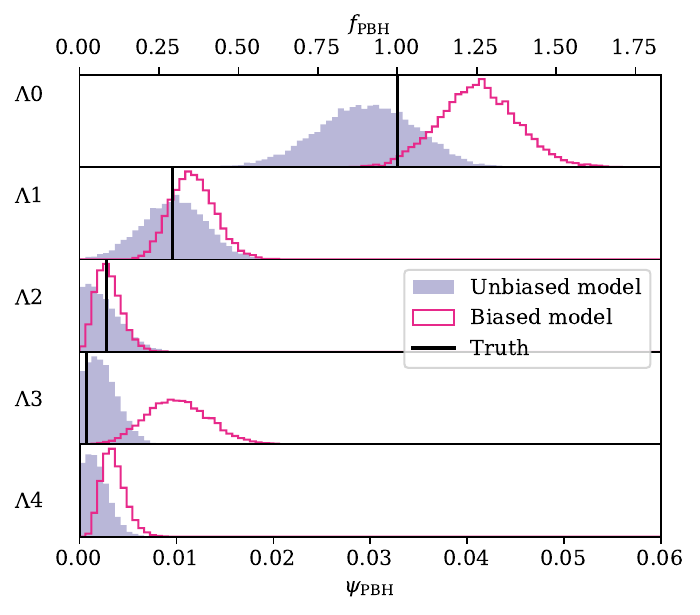}
\includegraphics[width=.48\linewidth]{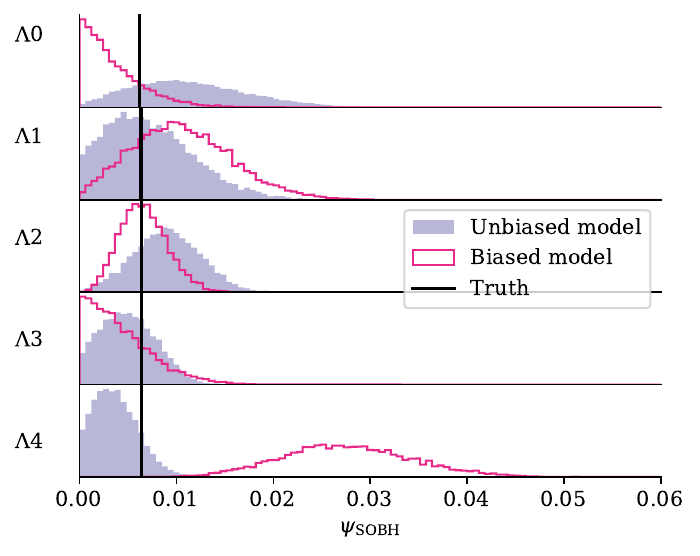}
\caption{
	\textbf{Top Row}: Posterior distributions on the relative abundance for the PBH (left) and SOBH (right) subpopulations, for both the full analysis and the restricted analysis plotted as the shaded region and the hatched region, respectively.
	The prior for the mixing fractions in the case of the restricted analysis was a broad Dirichlet distribution.
	There is good agreement between the two methods in the absence of systematic effects, as the true distributions of each subpopulation are assumed to be those of the injection for the restricted analysis. \textbf{Bottom Row}: Posterior distributions on the relative abundance for the PBH (left) and SOBH (right) subpopulation, for both the unbiased restricted analysis using the correct distribution (unhatched) and a systematically biased restricted analysis (hatched).
	The assumed mass distributions for the biased analysis are shown in Fig.~\ref{fig:massRidgeFixedModelSystematics}.
	The prior for the mixing fractions in the case of the restricted analysis was a broad Dirichlet distribution.
	While the conclusions are mildly consistent, a clear systematic bias can be seen in the posteriors.
}\label{fig:PopFractionRidgeFixedModelComparison}
\end{figure*}

\begin{figure}
\includegraphics[width=0.32\linewidth]{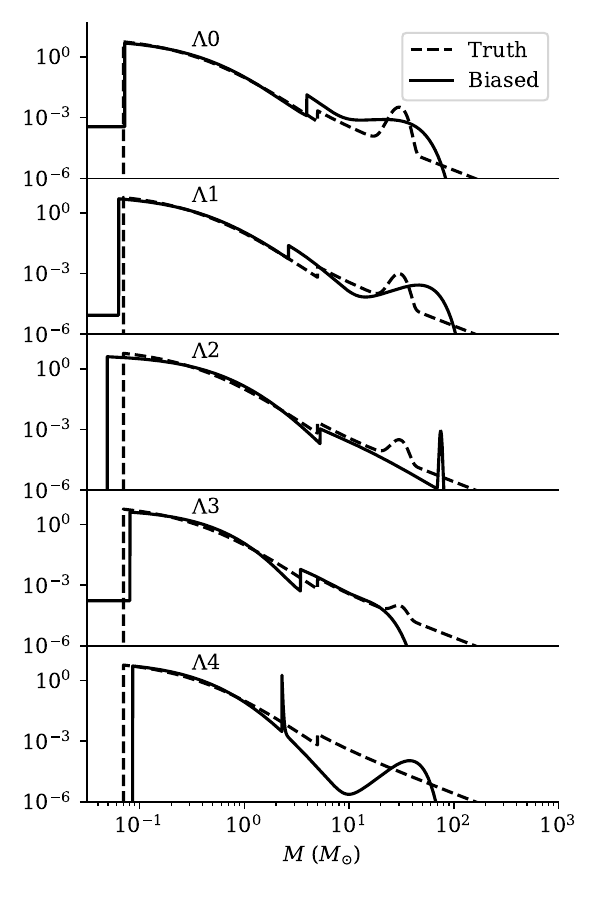}
\includegraphics[width=0.32\linewidth]{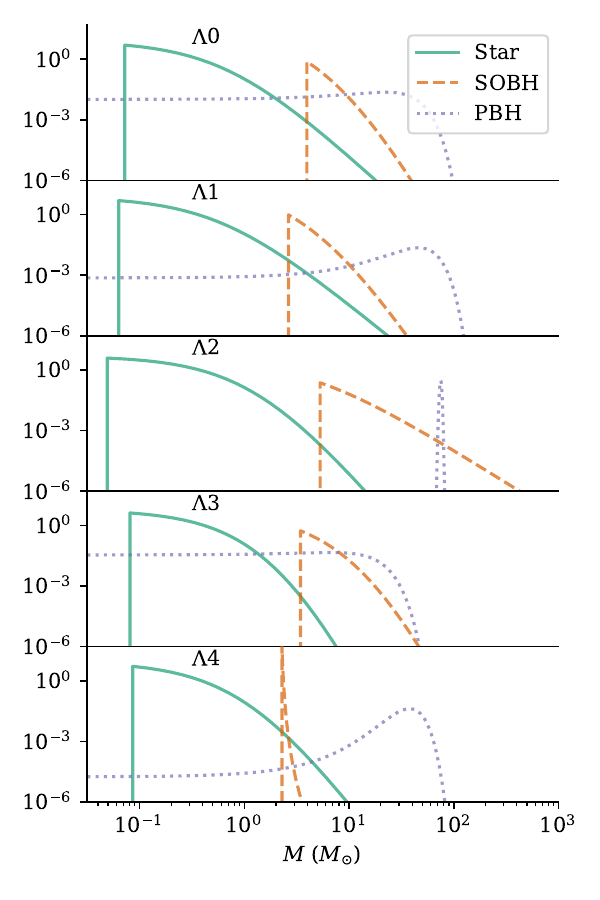}
\includegraphics[width=0.32\linewidth]{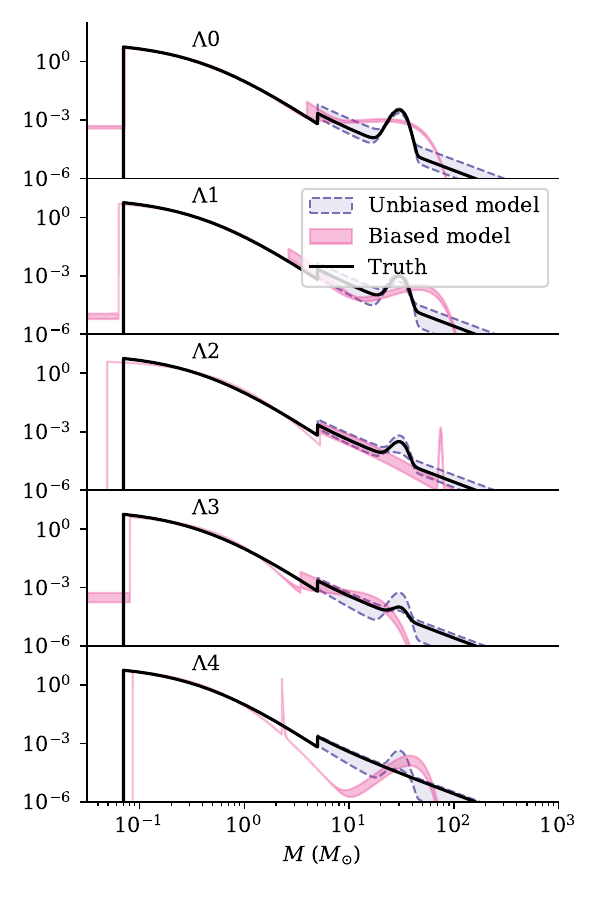}
\caption{
	\textbf{Left}: Mass distributions for the different data sets assumed while re-conducting the restricted analysis.
	The true distributions are the models used to generate the data (shown as a solid black line for each set of data), while the models used to re-apply the analysis are shown as a dashed black line.
    \textbf{Center}: The biased mass distributions for each subpopulation, where the stellar subpopulation is represented by a green solid line, the SOBH subpopulation is represented by a orange dashed line, and the PBH subpopulation is represented by a purple dotted line. 
    Each subpopulation distribution is normalized independently, so the relative heights do not reflect the total population.
	These mass distributions used to re-conduct this analysis were chosen by finding the $10$-th percentile of the posterior of the full analysis. 
    \textbf{Right:} The reconstruction of the mass distribution of the entire lens population from the posterior of the restricted analysis using the true underlying distributions (unbiased, non-hatched and green) and systematically biased assumptions (biased, hatched and orange). 
	The actual distributions are shown in black as dashed lines.
	We can see that the relative abundances favor biased values to compensate for systematics in the shapes of the distributions for each subpopulation.
}\label{fig:massRidgeFixedModelSystematics}
\end{figure}

We can now test this restricted analysis by using it to perform inference on the relative abundances and compare the results to the full analysis.
When sampling using the restricted analysis, we assign a prior consistent with the constraint $\sum_a \popfraca=1$. We will only sample $\fracA$ and $\fracB$ and use a Dirichlet prior informed by the original simulation ~\citep[e.g.,][]{Golovich:2020acu}. The Dirichlet distribution is parametrized by a vector $\boldsymbol{b}$,
\begin{equation}
\boldsymbol{b} =  (\hat{\popfrac}-{\rm min}(\hat{\popfrac}))  + c\,,
\end{equation}
where $c$ controls the width of the prior, and $\hat{\popfrac}$ is a vector of the relative abundances of the injected population model. We chose $c=1$ to produce a broad, uninformative prior on the parameters.

While more restricted in flexibility, this comes at the benefit of $\approx10^{2}$ computational speed up. The full population inference takes approximately 360 CPU hours on cluster-grade Intel Xeon E5-2695 nodes, and the restricted analysis takes approximately 1 CPU hour on a consumer-grade Intel Core i9-9980HK chip. Fig.~\ref{fig:PopFractionRidgeFixedModelComparison} (top row) shows the posterior constraints abundances of the PBH and SOBH subpopulations compared with the full population model used in Sec.~\ref{sec:populationInference}. We see good agreement between the two analyses, with constraints being tighter in the restricted model due its comparative lack of flexibility. This significantly more computationally efficient restricted analysis gives unbiased and similar constraints on the abundances to the full analysis when we condition on knowing the model that generated the data, however, it has to be validated in a more realistic setting before fully adopted. 

To investigate the sensitivity of the restricted analysis to biased assumptions about the underlying population, we take $10$-th percentile posterior sample from the full three subpopulation model for $\hyp_{0-4}$ and run the restricted analysis with an alternate assumption for the underlying form of the mass spectra. This procedure injects bias into our analysis which simulates not knowing the subpopulation models exactly. While any number of alternative mass spectra could have been picked, we chose distributions that were different from the truth while still being consistent with the data in $t_E$-$\pi_E$ space. Fig.~\ref{fig:massRidgeFixedModelSystematics} (left) shows the alternative mass spectra along the true spectrum used to generate the data.

Fig.~\ref{fig:PopFractionRidgeFixedModelComparison} (bottom row) shows the posterior constraints on the PBH and SOBH abundances using the restricted model, both assuming an unbiased and the new, biased population model. The general disagreement between these posteriors highlights the fact that biased modeling assumptions can impact the conclusions drawn with the restricted model. While the posteriors do overlap in each case, there is clear systematic bias, larger than the statistical error, that can lead to incorrect conclusions. Each of the biases can be connected with features in the assumed mass distribution. 
For example, for $\hyp_4$, we see that the stellar subpopulation in our assumed model is more peaked than the true distribution, leading to posteriors for both the PBH and SOBH relative abundances which are biased high, compensating for the lack of support at higher mass from the stellar subpopulation. Finally, Fig.~\ref{fig:massRidgeFixedModelSystematics} (right) shows the effect on the mass reconstruction assuming incorrect distributions for each subpopulation. The biased reconstructions favor biased mixing ratios to compensate for the systematics in the shape parameters between the true distribution and the assumed ones.

The inference from the restricted model suggests that it could provide a computationally fast and tractable alternative to the full model inference if the form of the population mass spectra is relatively well known \emph{a priori}. However, in the current realistic case that the population model is not well known, the restricted analysis could lead to biased conclusions. The full method of Sec.~\ref{sec:populationInference} can be extended to robustly marginalize over any reasonable population uncertainty and should be the default method for performing hierarchical inference until the lens population mass spectrum is better understood.

\section{Numerical Calculation of Population Efficiencies}\label{app:alphaCalc}


Examining Eq.~\eqref{eq:alpha}, we compute $\alpha$ by drawing samples from the population model and subsequently from the noise model under some simplifying assumptions. To match existing survey selection functions, we assume that $\pi_E$ has a small effect on selection allowing us to neglect it and only consider non-parallax parameters. This calculation can be computationally expensive relative to a likelihood evaluation, so we separate it into two steps. The first step being computing the integrals that do not change between population models we are exploring (such as the auxiliary parameters and the noise realizations), and secondly, computing the integrals that do change between our population models (namely, over $t_E$).
The first step only has be done once for the entire analysis.
With this categorization of parameters, Eq.~\eqref{eq:alpha} becomes
\begin{subequations}
\begin{align}\nonumber
\alpha &= \iiint p(\trig| \D)p( \D | \eparamnl,t_E) p( \eparamnl,t_E| \hyp)  d t_E  d\eparamnl d \D\,,\\
&= \iint p_{\rm det}(t_E) p( t_E| \hyp)d t_E  \,,
\end{align}
\begin{align}
p_{\rm det}(t_E) &\equiv \iint p(\trig| \D)p( \D | \eparamnl,t_E) p( \eparamnl| \hyp)d\eparamnl d\D  \,,
\end{align}
\end{subequations}
where $\eparamnl$ is the set of parameters excluding $t_E$ such that $\eparamnl = \eparam \,\backslash \,\{t_E\}$.
In the first equation, we have also assumed that the distribution of the parameters $\eparamnl$ are independent of $t_E$, which is true in our toy problem. 

We then construct a logarithmically spaced grid in $t_E$ that encapsulates the relevant parameter space ($0.1$ days $< t_E< 10^6$ days).
For each point in the grid ($t_{E,l}$), we calculate the partially-marginalized detection probability $p_{\rm det}(t_{E, l})$.
To do this, we draw event parameters from the population distribution $p(\eparamnl|\hyp)$, from which we calculate a light curve. 
We then perform the second integral over data realization through another Monte Carlo integral, where the noise is drawn from the likelihood $p(\D | \eparaml,t_{E,l})$.
Given our assumption that the noise is white and stationary we have,
\begin{align}\nonumber
p_{\rm det} (t_{E,l}) &\approx \frac{1}{S} \sum_j^S \left( \frac{1}{K} \sum_c^K p(\trig|\Dcj)\right) \,,\\\nonumber
\Dcj &\sim p(\D | {\eparamnl}_{,j}, t_{E,l})\,, \\
{\eparamnl}_{,j} &\sim p( \eparamnl| \hyp)\,.
\end{align}

\begin{figure}
\centering
\includegraphics[width=.5\linewidth]{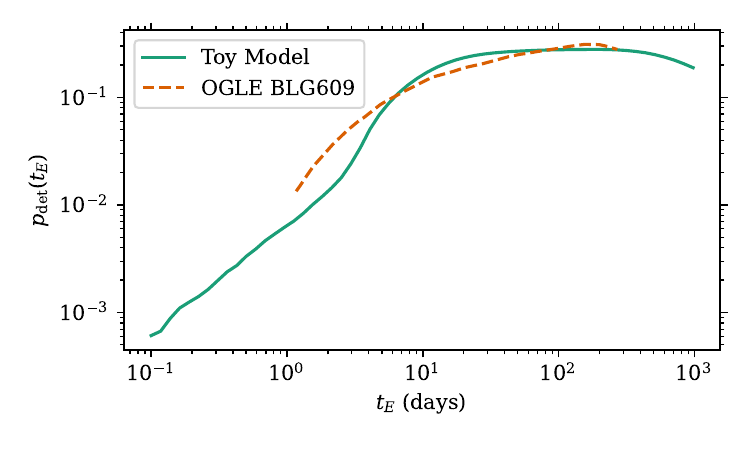}
\caption{
	The semi-marginalized detection probability curve, $p_{\rm det}(t_E)$ used in this exercise compared to the published OGLE-IV field BLG 609 detection efficiency curve.
}\label{fig:pdet_curve}
\end{figure}

$p_{\rm det}(t_E)$ calculated at each grid point, we can interpolate across the detection probability values, giving the typical detection efficiency curves published by many surveys.
The results of this calculation are shown in Fig.~\ref{fig:pdet_curve}, compared to a published detection probability curve for a specific field from OGLE for reference.
With this interpolated function, we can now finish the rest of the detection efficiency calculation repeated for every population model we consider,
\begin{align} \nonumber
\alpha &= \int d \D \int d \eparam p(\trig| \D) p(\D | \eparam) p(\eparam| \hyp)\,, \\ 
&= \int d t_E p_{\rm det} (t_E) p(t_E | \hyp) \,.
\end{align} 
This is now a one dimensional integral which can estimated using,
\begin{subequations}
\begin{equation}
\alpha \sim \frac{1}{S} \sum_{c = 0}^S  p_{\rm det} (t_{E,c}) \,,
\end{equation}
\begin{equation}
t_{E,c} \sim p(t_E | \hyp)\,, 
\end{equation}
\end{subequations}
for $S$ independent samples.
When considering the restricted hierarchical model, this can also be done for each subpopulation, independently.


\bibliography{refs}{}
\bibliographystyle{aasjournal}
\end{document}